\documentclass[11pt]{article}

\usepackage{fontspec}
\usepackage{polyglossia}

\usepackage{subcaption}
\usepackage[preprint]{acl}

\usepackage{times}
\usepackage{tikz}
\usepackage[edges]{forest}
\usepackage{graphicx}
\usepackage{times}
\usepackage{latexsym}
\usepackage{times}
\usepackage{latexsym}
\usepackage{booktabs}
\usepackage{boxedminipage}
\usepackage{amssymb}
\usepackage{amsmath}

\usepackage{graphicx}
\usepackage{enumitem}
\usepackage{listings}
\usepackage{xstring}
\usepackage{subcaption}
\usepackage{latexsym}
\usepackage{amsmath}
\usepackage{tcolorbox}
\usepackage{longtable}  
\usepackage{booktabs}   
\usepackage{array}      
\usepackage{geometry}   
\usepackage{booktabs}   
\usepackage{tabularx}   
\usepackage{multirow}
\usepackage{caption}
\usepackage{ragged2e}
\usepackage[T1]{fontenc}

\usepackage[utf8]{inputenc}

\usepackage{microtype}

\usepackage{inconsolata}

\usepackage{graphicx}
\usepackage{tikz}
\usepackage{forest}
\usepackage{pdflscape}  
\usepackage{booktabs}
\usepackage{tabularx}
\usepackage{array}
\usepackage{ragged2e}
\usepackage{makecell}

\usepackage{multirow}
\usepackage{array}
\usepackage{ragged2e}
\usepackage{colortbl}
\usepackage{algorithm}
\usepackage{algorithmic}

\newcolumntype{P}[1]{>{\RaggedRight\arraybackslash}p{#1}} 
\newcolumntype{R}[1]{>{\raggedleft\arraybackslash}p{#1}}  

\definecolor{RowSM}{RGB}{235,245,255}      
\definecolor{RowTPC}{RGB}{235,255,240}     
\definecolor{RowADS}{RGB}{255,245,235}     
\definecolor{RowTPCADS}{RGB}{245,235,255}  

\definecolor{RowC1}{RGB}{242,248,255} 
\definecolor{RowC2}{RGB}{228,241,255}
\definecolor{RowC3}{RGB}{214,234,255}
\definecolor{RowC4}{RGB}{200,227,255} 

\usepackage{graphicx}
\definecolor{lightblue}{rgb}{.50,.95,1}
\definecolor{tri}{rgb}{.25,.88,.82}
\definecolor{lilac}{rgb}{0.85,0.64,0.85}

\usepackage[table]{xcolor}
\definecolor{phasecolor}{RGB}{230,223,250}
\definecolor{totalcolor}{RGB}{226,232,240}
\definecolor{rcolor}{RGB}{230,223,200}

\lstdefinelanguage{prompt}{
  sensitive=true,
  morecomment=[l]{System},
  morecomment=[l]{ASR},
  morecomment=[l]{Dialect},
  morecomment=[l]{Emotion},
  morecomment=[l]{SSUM},
  morecomment=[l]{TSUM},
}

\usepackage{url}
\usepackage[colorinlistoftodos,prependcaption,textsize=small]{todonotes}
\presetkeys{todonotes}{inline}{}

\usepackage[table]{xcolor}
\definecolor{genlight}{HTML}{E8F5E9}   
\definecolor{disclight}{HTML}{E3F2FD}  

\definecolor{genlight}{RGB}{220,240,220}
\definecolor{disclight}{RGB}{210,225,245}
\newcommand{\dgain}[1]{\textcolor{teal}{\textbf{#1}}}
\newcommand{\dloss}[1]{\textcolor{orange!90!red}{#1}}

\title{Multi-Task Instruction Tuning via Data Scheduling \\ for Low-Resource Arabic SpeechLLMs}

\author{
Hunzalah Hassan Bhatti,
Firoj Alam, 
Shammur Absar Chowdhury \\
Qatar Computing Research Institute, Qatar \\
hunzalahhassan@gmail.com, \{fialam, shchowdhury\}@hbku.edu.qa \\
}

\setmainlanguage{english}
\setotherlanguage{arabic}

\begin{document}
\maketitle

\begin{abstract}
Audio large language models (LLMs) enable unified speech understanding and generation, but adapting them to linguistically complex and dialect-rich settings such as Arabic--English remains challenging. We present a controlled study of multi-task instruction tuning for an Arabic-centric audio LLM across generative tasks including ASR and speech and text summarization, and discriminative tasks including dialect and emotion recognition, in a resource-constrained setting. To support end-to-end Arabic speech summarization, we introduce \textbf{AraMega-SSum}, a \textit{first} speech summarization resource for training and benchmarking Arabic-centric Audio-LLMs. We compare four training strategies \textit{(i)} \textit{Uniform Mixing (UM)}, \textit{(ii)} \textit{Task-Progressive Curriculum (TPC)}, \textit{(iii)} \textit{Aligner-Based Diverse Sampling (ADS)} for training-time batch construction, and \textit{(iv)} A two-stage \textit{TPC$\rightarrow$ADS} strategy. 
Our results reveal a clear \textit{efficiency--robustness trade-off}. TPC yields the
strongest generative performance on ASR and summarization. ADS improves paralinguistic
tasks but reduces generative stability when used alone. The two-stage
\textit{TPC$\rightarrow$ADS} strategy provides the best overall balance, delivering
the strongest DID and SER results while outperforming large proprietary models such as
Gemini-2.5-Pro on discriminative tasks. We will make \textbf{AraMega-SSum} and all
experimental resources publicly available to the
community.\footnote{\url{anonymous.com}}
\end{abstract}

\section{Introduction}
\label{sec:intro}

Large language models (LLMs) are rapidly evolving into native multimodal systems capable of unified speech understanding and generation \cite{xu2025qwen25omnitechnicalreport}. However, adapting these ``omni'' models to linguistically complex and low-resource settings such as the Arabic--English bilingual space remains challenging. This setting combines dialectal variation, 
code-switching, and paralinguistic cues such as emotion, which are often difficult for general-purpose audio backbones to model effectively. 
Instruction tuning for this setting should jointly handle \textit{generative} tasks such as ASR and summarization, and \textit{discriminative} tasks such as dialect and emotion recognition. In practice, this is further complicated by severe imbalance across tasks and labels. Large ASR corpora can dominate optimization, while smaller paralinguistic datasets and minority classes may remain under-trained, resulting in negative transfer and reduced robustness \citep{chen2024octavius}.

We study how \textbf{data scheduling} and \textbf{within-batch task mixing} affect transfer, training stability, and final performance. \textit{Data scheduling} controls how tasks and examples are sampled over training time, whereas \textit{within-batch mixing} controls which tasks, labels, and acoustic conditions co-occur in the same batch.

Our main experiments use \textbf{Qwen2.5-Omni-7B}~\cite{xu2025qwen25omnitechnicalreport}, a dense open-weight omni model and our primary Arabic-centric adaptation target. We additionally include \textbf{Gemma-4-E4B-it}\footnote{\url{https://huggingface.co/google/gemma-4-E4B-it}} as a secondary contrastive model to test whether the observed scheduling trends persist under a different architecture with stronger Arabic priors, including prior Arabic ASR exposure. Thus, Qwen captures the main low-resource Arabic adaptation setting, while Gemma serves as an architecture-sensitivity check.

To support end-to-end Arabic speech summarization, we introduce \textbf{AraMega-SSum}, a \textit{first} Arabic speech summarization dataset for training and benchmarking Arabic-centric AudioLLMs. It contains 49,995 training samples and 4,000 test samples, corresponding to 222 and 16.98 hours of speech, respectively. The dataset is built from MSA-translated summaries and speaker-conditioned synthesized speech from 47 speakers across 10 Arab countries. A subset of the data is human-evaluated. AraMega-SSum is used only for summarization, while ASR, dialect identification, and emotion recognition are evaluated on real-speech benchmarks.

\begin{figure*}[th]
    \centering
    \includegraphics[width=0.9\linewidth]{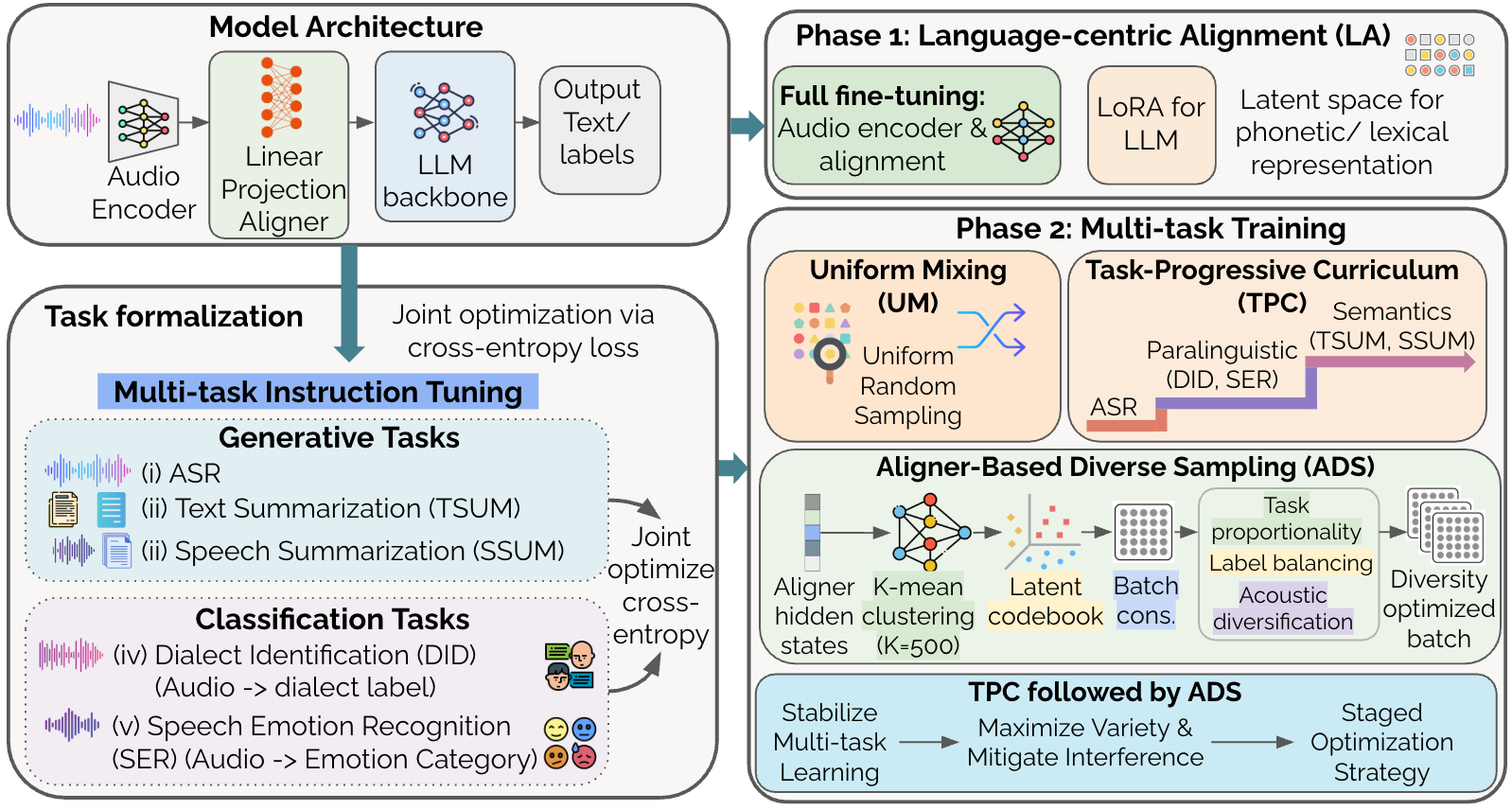}
    \vspace{-0.3cm}
    \caption{Overview of the proposed methodology. The framework adapts two AudioLLM backbones for a unified set of generative and classification audio tasks. The Qwen-based model uses a Whisper-v3 audio encoder with a Qwen2.5-7B LLM backbone, while the Gemma-based model uses a USM-style encoder with a Gemma-4 LLM backbone. Training proceeds in two stages: language-centric alignment (Phase 1), followed by multi-task adaptation (Phase 2), where we compare UM, TPC, ADS, and the two-stage TPC-ADS strategy.
    }
    \label{fig:methodology}
    \vspace{-0.35cm}
\end{figure*}

We instruction-tune Qwen2.5-Omni-7B and Gemma-4-E4B-it with LoRA and compare four training strategies. \textit{(i)} \textbf{Uniform Mixing (UM)} is the standard baseline, sampling uniformly from the pooled multi-task data. \textit{(ii)} \textbf{Task-Progressive Curriculum (TPC)} introduces tasks gradually, moving from lower-level acoustic objectives to higher-level objectives. \textit{(iii)} \textbf{Aligner-Based Diverse Sampling (ADS)} preserves task proportions, balances labels for classification tasks, and \textit{promotes cluster diversity} in the \textbf{aligner embedding space} that maps speech encoder representations to the LLM. \textit{(iv)} \textbf{TPC$\rightarrow$ADS} combines early-stage curriculum stabilization with later diversity-oriented training.


Our contributions are the following:
\begin{itemize}[noitemsep,topsep=0pt,labelsep=.5em] 
    \item We provide a controlled and comparative study of scheduling and batch composition for Arabic-centric omni audio instruction tuning across Generative tasks: \textit{ASR and summarization}, and Discriminative tasks: \textit{dialect identification, and emotion recognition}. 
 \item We introduce \textbf{AraMega-SSum}, a novel and first dataset for Arabic speech summarization, with human evaluation on a subset of the dataset.
    \item We show a consistent efficiency--robustness trade-off across training schedules. \textbf{TPC$\rightarrow$ADS} provides a reliable balance for discriminative tasks with imbalanced classes, whereas \textbf{TPC} is more effective for generative tasks. 
\end{itemize}

To the best of our knowledge, this is the \textit{first} work to study specialized training strategies for an \textbf{Arabic-centric omni-model} spanning acoustic, linguistic, and paralinguistic tasks. It is also the \textit{first} study to show how diversity in aligner embeddings can guide training-time batch construction while promoting task balance, label balance, and acoustic diversity.
We emphasize that our contribution is not generic multi-task data mixing, but a controlled study of how task scheduling and batch construction affect Arabic-centric AudioLLM adaptation under limited resource.

Our findings reveal a consistent trade-off across training schedules, with variation 
across model families. For \textbf{Qwen2.5-Omni-7B}, TPC gives the best ASR and 
matches UM on summarization, confirming that curriculum ordering stabilizes 
acoustic-to-text mapping. For \textbf{Gemma-4-E4B-it}, UM gives the best ASR, 
likely because its multilingual pretraining already provides a strong Arabic prior, 
reducing sensitivity to curriculum ordering. For discriminative tasks, 
TPC$\rightarrow$ADS gives the best DID for both models and the best SER for 
Qwen2.5-Omni-7B, with a notable $+$16.6 gain over TPC, showing that 
diversity-oriented sampling is most effective for paralinguistic tasks once the 
shared representation has been stabilized. Overall, \textbf{TPC} is preferable for 
generative tasks, while \textbf{TPC$\rightarrow$ADS} provides the strongest balance 
for discriminative tasks with imbalanced classes.

\section{Methodology}
\label{sec:methodology}

In Figure \ref{fig:methodology}, we present an overview of the proposed methodology. We study data scheduling and within-batch task mixing for adapting an omni audio LLM to Arabic-centric speech understanding. We utilized the \textit{Qwen2.5-7B-Omni} and \textit{Gemma-4-E4B-it} architecture. For Qwen2.5-7B-Omni combines a Whisper-v3 audio encoder \cite{radford2023robust} with a 7B transformer decoder. A linear projection layer, called 
the \textbf{aligner}, maps encoder features into the LLM embedding space. We did the same for Gemma-4-E4B-it with USM-style audio encoder~\cite{zhang2023google}.

\subsection{Tasks Formalization}
We instruction-tune the model on a unified task, $\mathcal{T} = \{\mathcal{T}_{g}, \mathcal{T}_{c}\}$, set that includes generative ($\mathcal{G}$) tasks  and classification ($\mathcal{C}$) tasks. Each training example consists of an audio input $\mathbf{x}$ (or a transcript for text-only summarization), a task prompt $\mathbf{p}$, and a target output sequence $\mathbf{y}$. We optimize the negative log-likelihood of the target tokens using cross-entropy.

\noindent
\textbf{Generative ($\mathcal{G}$) tasks}
We include automatic speech recognition (ASR), text summarization (TSUM) given the transcript, and speech summarization (SSUM) given the speech input. For TSUM and SSUM, the target is a short summary in Arabic or English depending on the prompt.

\noindent
\textbf{Classification ($\mathcal{C}$) tasks}
We include dialect identification (DID) and speech emotion recognition (SER). We formulate classification in the same instruction format by predicting a canonical label string, so the loss remains token-level cross-entropy.

\subsection{Two-phase Adaptation}
We use two training phases to separate modality alignment from multi-task scheduling effects.


\noindent
\textbf{Phase 1: Language-centric alignment.}
We fine-tune the audio encoder and aligner on large-scale bilingual ASR to align the audio representation with Arabic and English phonetic and lexical structure. During this phase, the LLM is adapted using LoRA.

\noindent
\textbf{Phase 2: Multi-task instruction tuning.}
We freeze the audio encoder and the aligner, and we update only the LoRA parameters in the LLM. All scheduling strategies in this phase use the same compute budget, meaning the same total number of training steps, so differences are attributable to the sampling strategy.

\subsection{Data Scheduling and Batch Construction}


In this section, we compare four strategies under the same model, optimization settings, and fixed computational setup. 

\paragraph{A. Uniform Mixing Baseline (UM):} We sample training instances uniformly from the pooled multi-task set $\mathcal{D}$. This is a standard baseline for instruction tuning.

\paragraph{B. Task-Progressive Curriculum (TPC):} TPC trains in different stages ordered by task abstraction. The model is exposed to tasks in five sequential stages: \textcolor{blue}{\textbf{Acoustic}}: ASR $\rightarrow$ \textcolor{blue}{\textbf{Paralinguistic}}: DID, SER $\rightarrow$ \textcolor{blue}{\textbf{Semantic abstraction}}: TSUM, SSUM.
At each stage, we retain a fixed fraction of data from earlier stages to reduce forgetting.

\paragraph{C. Aligner-Based Diverse Sampling (ADS):} 
To mitigate task interference and label imbalance, particularly the dominance of MSA over regional dialects such as Gulf Arabic, we propose the ADS method, presented in Algorithm \ref{alg:ads}. The method promotes batch diversity by sampling from a discretized latent space obtained through K-means clustering of the aligner's output representations.

\begin{algorithm}[t]
\footnotesize
\caption{ADS batch construction}
\label{alg:ads}
\begin{algorithmic}[1]
\STATE $h_i \leftarrow A_\phi(x_i)$ for all $x_i \in \mathcal{D}$
\STATE $\mathcal{C} \leftarrow \mathrm{KMeans}(\{h_i\}, K=500)$
\STATE Assign each $x_i$ to nearest centroid $c_k \in \mathcal{C}$
\WHILE{training}
    \STATE $\mathcal{B} \leftarrow \emptyset$
    \FOR{each task $t \in \mathcal{T}$}
        \STATE $n_t \leftarrow M \cdot \mathrm{PriorDist}(t)$
        \FOR{each label $l \in \mathrm{UniqueLabels}(t)$}
            \STATE $\mathcal{S}_{t,l} \leftarrow \{x_i : \tau_i=t,\, l_i=l\}$
            \STATE Sample $x \sim \mathcal{S}_{t,l}$ by round-robin over clusters
            \STATE $\mathcal{B} \leftarrow \mathcal{B} \cup \mathrm{UpsampleMinor}(x)$
        \ENDFOR
    \ENDFOR
    \STATE \textbf{yield} $\mathcal{B}$
\ENDWHILE
\end{algorithmic}
\end{algorithm}
\noindent \textbf{Latent representation.} We first define a representation space using the hidden states of the linear aligner, $A_\phi$. Given that the aligner bridges the acoustic encoder and the LLM, its embeddings capture the specific phonetic-semantic features relevant for downstream semantic understanding. To minimize computational overhead, we perform clustering on a 3\% representative subset of the total corpus. For each sample $i$, we extract the aligner hidden states and apply max-pooling over the temporal dimension to obtain a fixed-size vector $h_i \in \mathbb{R}^d$. We then apply $K$-Means clustering to generate a global acoustic-semantic codebook $\mathcal{C}$ with $K=500$ centroids.

\noindent \textbf{Batch construction and upsampling.}
During the SFT phase, each ``effective batch'' $\mathcal{B}$ is constructed to satisfy three constraints:
\begin{enumerate}[noitemsep,topsep=0pt,labelsep=.5em] 
    \item \textit{Task proportionality:} The relative frequency of tasks in batch $\mathcal{B}$ follows the original distribution of the dataset to maintain stable convergence on high-volume tasks like ASR.
    \item \textit{Label balancing:} For discriminative tasks (e.g., Emotion, Dialect), we upsample minority classes ($UpsampleMinor(.)$) such that all labels, $l$, within a task appear with prior frequency per batch.
    \item \textit{Acoustic diversification:} Samples are selected using a Round-Robin traversal across the $K$ clusters within each task-label pair. This ensures that the model is exposed to a maximally diverse set of speakers and acoustic environments in every gradient update.
\end{enumerate}



\paragraph{D. TPC followed by ADS:}
We also evaluate a two-stage training strategy that runs TPC for an initial portion of training to stabilize the shared representation, then switches to ADS for the remaining steps to emphasize label coverage and cluster diversity. This strategy uses the same total number of steps as the other strategies.

\section{Datasets}
\label{sec:datasets}
To facilitate our language-centric alignment and multi-task training objectives, we curate training and evaluation data covering MSA, multiple Arabic dialects, and English. The development of Arabic-centric audio LLMs is challenging due to dataset scarcity. Large-scale ASR data are relatively available, whereas high-level spoken understanding and paralinguistic tasks remain limited and often exhibit strong label imbalance. A major gap is Arabic end-to-end speech summarization, for which paired speech and abstractive summaries are scarce in the public domain. AraMega-SSum is the only newly developed semi-synthetic dataset in our study, introduced to support this missing Arabic speech summarization setting. The remaining datasets for ASR, emotion recognition, and dialect identification are curated from public real-speech benchmarks.

\subsection{AraMega-SSum}
To address current gaps in the literature, we introduce and publicly release \textbf{AraMega-SSum}, a large-scale Arabic speech summarization dataset with 49,995 training samples (222 hours) and 4,000 test samples (16.98 hours). AraMega-SSum (Arabic Short Speech Summarization) serves as the first end-to-end dataset for semantic \emph{understanding} directly from Arabic audio, providing a foundation for evaluating high-level generative understanding in Arabic-centric multimodal systems. A key contribution is the dataset's scale and quality for Arabic spoken short summarization. Given the extreme scarcity of paired audio-to-summary data in Arabic, we construct AraMega-SSum using a semi-synthetic pipeline that preserves semantic consistency while maximizing acoustic diversity. Our goal is to provide a large dataset that enables reproducible Arabic speech summarization evaluation, together with quality controls that reduce noise.

\noindent
\textbf{Translation.}
We leverage the \textbf{MegaSSUM} corpus \cite{matsuura24_interspeech} as our source, a large-scale English sentence-wise speech summarisation dataset based on the Gigaword dataset, comprising more than 3.8 million synthesized speech, transcription, and summary triplets. English speech is generated using a multi-speaker text-to-speech model trained on LibriTTS-R to produce natural-sounding utterances from the first sentences of news articles paired with their headlines, enabling broad coverage and consistent speech-text alignment. We translated the English source transcripts and their corresponding human-written summaries into MSA using \textit{Gemini-2.5-flash}. 

\noindent\textbf{Translation Quality.} To ensure the naturalness of the translations, we utilise GPT-4.1 as judge to evaluate the quality of translations with respect to \textit{semantic equivalence}, \textit{information preservation}, \textit{contextual accuracy}, \textit{completeness}, and \textit{coherence}. The prompt is provided in Appendix Listing~\ref{fig:llm_translation_judge_prompt}. Appendix Table~\ref{tab:translation_quality} reports LLM-as-a-judge scores on a 0--10 scale. 
Moreover, we also performed a human evaluation to validate the translation quality. Table~\ref{tab:translation_quality_hh} shows a \textit{similar pattern} based on \textbf{\textit{human evaluation}} of a subset of 200 pairs. Under our current setup and rubric, most outputs are fluent and semantically aligned with the source.




\noindent
\textbf{Audio Synthesis via Neural Voice Cloning.}
To reconstruct the speech modality, we use \textbf{XTTS-v2} \cite{casanova2024xtts}, a state-of-the-art multi-speaker latent diffusion model for speech synthesis. 
To reduce the risk of overfitting to static synthetic acoustic profiles, we leverage the model's zero-shot \textbf{voice cloning} capability. We use reference speaker audio from \textit{MenaSpeechBank}~\cite{ali2026menaspeechbankreferencevoicebank} and extract speaker embeddings from \textbf{47 distinct speakers} representing \textbf{10 Arab countries}.
This geographic breadth of the speakers audios helps ensure that the generated speech captures a wide range of regional prosody, pitch variation, and glottal characteristics across the Arab world. For each reference speaker, we maintain at least 10 high-quality voice clips, and randomly sample one reference clip to increase intra-speaker acoustic variability. This approach provides the necessary acoustic diversity for the LLM to generalize across heterogeneous recording conditions.
The resulting dataset, \textbf{AraMega-SSum}, contains $\approx$\textbf{50K} pairs and provides a strong foundation for learning semantic representation.

\noindent\textbf{Audio Quality.}
We validated the generated speech using both human assessment and standard automatic quality metrics. Three paid annotators rated 100 randomly sampled test-set clips for \textit{naturalness}, \textit{audio quality}, and \textit{intelligibility}~\cite{10.5555/3692070.3692979,du2025cosyvoice}. The clips achieved average scores of 3.98, 4.37, and 3.69 out of 5, respectively, with substantial-to-near-perfect annotator agreement (78.9 to 95.0) measured by Gwet's AC1~\cite{gwet2008computing} (Appendix Table \ref{tab:human-speech-quality}). Complementary metrics show that the generated speech is generally intelligible and preserves speaker characteristics: Fanar Aura ASR obtains a WER of $15.5$,
NISQA gives an overall score of $3.95$ out of 5
and speaker cosine similarity with the reference speakers is $0.66$.
\subsection{Training Datasets}

\paragraph{Datasets for language-centric alignment.} 
For the model alignment experiments, we curate a high-quality ASR dataset of $\approx$\textbf{15K hours} from multiple sources, as shown in Appendix Table~\ref{tab:dataset_stats}. We select these sources to ensure balanced linguistic coverage while maintaining computational efficiency. To improve acoustic robustness under real-world variability, we apply a targeted data augmentation strategy. Specifically, we  apply \textbf{speed perturbation} (with factors of 0.9$\times$ and 1.1$\times$), together with \textbf{additive noise augmentation} using the MUSAN \cite{musan2015} and RIR \cite{ko2017rir} corpora on a subset. This process yields an additional $\approx$15,000 hours of synthetic high-variance audio.
The final dataset consists of $\approx$\textbf{27K hours}.


\paragraph{Datasets for multi-task training.}
For the multi-task learning experiments, we selected several datasets covering both generative and discriminative tasks, including ASR, summarization, emotion recognition, and dialect identification. Below, we briefly describe each dataset.

\begin{itemize}[leftmargin=*,noitemsep,topsep=0pt,labelsep=.5em] 
    \item \textbf{ASR:} Given the high-resource nature of ASR compared to downstream semantic tasks, we performed a controlled \textbf{subsampling of the ASR dataset}. We selected 6.44\% of the dataset curated for Phase 1. 
    \item \textbf{Summarization:} For English, we use the standard training partitions of the Mega-SSum dataset. For Arabic, we use the newly developed \textbf{AraMega-SSum} dataset. Examples from the dataset are presented in Appendix Figure \ref{fig:ssum_eg}.
    \item \textbf{Emotion Recognition:} We curated publicly available Arabic affective corpora, which covers  spontaneous emotional prosody.
    \item \textbf{Dialect Identification (DID):} We unified the ADI-17 \cite{fan2020adi17} corpus with the MSA partitions from the ADI-5 dataset, providing a robust taxonomy covering six regional variants. 
\end{itemize}

Our final multi-task dataset aggregates approximately \textbf{5,109} hours of speech. To ensure training stability and optimize memory utilization, we constrained the duration of all speech instances to $t \leq 180$ seconds. Table \ref{tab:dataset_stats}, in the Appendix,
   reports the statistics of the curated datasets. Per-task, label-wise distributions are shown in Figures \ref{fig:did_disi}--\ref{fig:emotion_train_distribution} in the Appendix.

\subsection{Evaluation Datasets}
\label{sec:test_datasets}

To assess the multi-task capabilities we curate a comprehensive evaluation suite spanning generative and discriminative tasks from publicly available test sets. Below we describe each of them as in also reported in Appendix Table \ref{tab:benchmarks}.


\noindent\textbf{ASR.} 
We benchmark transcription performance across three dimensions: native English, L2 speech, MSA and dialectal Arabic. For native English, we utilize the \textbf{LibriSpeech} \cite{panayotov2015librispeech} \textit{test-clean} and \textit{test-other} partitions. 
To assess the model's proficiency with non-native accents, we evaluate on the \textbf{L2-ARCTIC} test split \cite{zhao2018l2arctic}, specifically filtering for speakers with Arabic as their first language (L1) to measure L2-English phonetic adaptation.
For Arabic, we employ the widely recognized \textbf{MGB-2} benchmark \cite{ali2016mgb2} for Modern Standard Arabic (MSA). We further evaluate the model’s robustness to dialectal variation and code-switching through a suite of challenging datasets:
\begin{itemize}[leftmargin=*,noitemsep,topsep=0pt,labelsep=.5em] 
    \item \textbf{Dialectal Diversity:} We use the \textbf{MGB-3} Egyptian test set \cite{ali2017mgb3} and the \textbf{SADA} \cite{alharbi2024sada} Saudi dialectal corpus to measure regional linguistic adaptation;
    \item \textbf{Code-Switching (CS):} We evaluate intra-sentential and inter-sentential dynamics using \textbf{ESCWA} (Arabic-English CS) \cite{chowdhury_towards_2021} and \textbf{DACS} (within-dialect CS) \cite{chowdhury2020dacs}.
\end{itemize}

Overall, this setup supports broad evaluation across English and Arabic, including native, non-native, dialectal, and code-switched speech.

\noindent\textbf{Emotion Recognition (ER).}
We evaluate emotion recognition on  \textbf{KSU Emotions:} acted Arabic speech \cite{alotaibi2020ksu}


\noindent\textbf{Dialect Identification (DID).}
We use the \textbf{ADI17} test set \cite{fan2020adi17} to evaluate classification across 17 Arabic dialects.
This serves as a high-granularity evaluation of the model's ability to distinguish subtle phonetic and lexical variations across the Arab world.

\noindent\textbf{Summarization (SSum and TSum).}
We evaluate summarization using the test sets of \textbf{Mega-SSum} for English \cite{matsuura24_interspeech} and our proposed \textbf{AraMega-SSum} for Arabic. These benchmarks measure the model’s ability to generate concise, faithful summaries from short-form audio. To separate acoustic errors from summarization ability, we also include a \textit{text-only} setting that feeds gold transcripts directly to the LLM, providing an oracle upper bound on summarization performance.

\section{Experimental Setup}
\label{sec:setup}

\paragraph*{Baseline.}
For baseline comparisons, we include both open and proprietary models. As an open baseline, we evaluate Qwen2.5-Omni-3B. For proprietary baselines, we compare against GPT-audio, GPT-5-chat, and Gemini-2.5-Pro, all evaluated in zero-shot mode without any task-specific adaptation.

\paragraph*{Two-Phase Training.}
We train in two phases and keep all hyperparameters fixed across all training strategies so differences are attributable to scheduling and batch construction.

\begin{itemize}[noitemsep,topsep=0pt,parsep=0pt,partopsep=0pt,leftmargin=*,labelsep=.5em]
\item \textit{Phase 1 Language-centric Alignment (LA).}
We perform foundational alignment for 1 epoch using the ASR task. In this phase, the audio encoder and linear aligner are fully fine-tuned while for the LLM, we adapted LoRA based training.
\item \textit{Phase 2 Multi-task instruction tuning} We run the four training strategies, UM, TPC, ADS, and TPC+ADS, for $\approx$10K steps. During this stage, the encoder and aligner remain frozen, and only the LoRA adapters in the LLM are updated. 

For ADS, we build a global codebook with $K=500$ clusters from a 3\% subset ($\approx 75$ hours) of the training data to capture fine-grained paralinguistic variation.
\end{itemize}

\paragraph*{Evaluation}
We evaluate generative and discriminative tasks with standard metrics. For \textbf{ASR}, we report WER, applying Arabic normalization by unifying Alef/Hamza variants and removing diacritics. For \textbf{summarization}, we report ROUGE-L~\cite{lin2004rouge}, BERTScore \cite{zhang2019bertscore}, and GPT-4.1 as an LLM judge \cite{zheng2023judging}. The judge scores outputs from 1--10 on clarity, conciseness, coherence, completeness, semantic alignment, accuracy, relevance, and information density (see prompts used in {Figure~\ref{fig:llm_judge_prompt}}). For \textbf{classification (SER and DID)}, we use weighted F1 and normalize label variants (e.g., mapping ``KSA'' to ``Saudi Arabia'') during post-processing.


\section{Results and Discussion}
\label{sec-results}

\subsection{Results Across Tasks}
In Table~\ref{tab:model-specific-phase-summary}, we report results, averaged over testsets, across adaptation phases. Phase 2 multi-task training improves both models over the base and LA settings. The gains are clearest for DID, SER, and summarization. 


Appendix~\ref{sec_app_additional_results} provides the full per-dataset breakdown. Tables~\ref{tab:asr-detailed-wer}--\ref{tab:did-detailed} report detailed results for ASR, SSUM and TSUM, SER, and DID. Table \ref{tab:ssum-llm-judge} includes LLM-as-judge results for summary tasks.

\begin{table}[!tbh]
\centering
\setlength{\tabcolsep}{3.2pt}
\scalebox{0.6}{
\begin{tabular}{lrrrr}
\toprule
\textbf{Phase / Model} 
& \multicolumn{2}{c}{\textbf{Generative}} 
& \multicolumn{2}{c}{\textbf{Discriminative}} \\
\cmidrule(lr){2-3} \cmidrule(lr){4-5}
& \textbf{ASR} $\downarrow$ 
& \textbf{Summ.} $\uparrow$ 
& \textbf{DID} $\uparrow$ 
& \textbf{SER} $\uparrow$ \\
\midrule
\multicolumn{5}{c}{\textbf{Qwen2.5-Omni-7B}} \\
\midrule
Base        & 48.79             & 55.0             & 10.3             & 13.71          \\
LA          & 21.70             & 54.3             & 9.8              & 24.43          \\
UM          & \underline{19.01} & \textbf{66.9}    & \underline{87.8} & 59.99          \\
TPC         & \textbf{18.97}    & \textbf{66.9}    & \underline{87.8} & 59.39          \\
ADS         & 19.60             & 64.9             & 86.7             & \underline{72.01} \\
TPC$\to$ADS & 19.04             & \underline{66.6} & \textbf{88.8}    & \textbf{75.94} \\
\midrule
\rowcolor{gray!8}
$\Delta$ TPC vs.\ UM
  & \dgain{$-$0.04} & {0.0} & {0.0} & \dloss{$-$0.6} \\
\rowcolor{gray!8}
$\Delta$ TPC$\to$ADS vs.\ TPC
  & \dloss{$+$0.07} & \dloss{$-$0.3} & \dgain{$+$1.0} & \dgain{$+$16.6} \\
\midrule
\multicolumn{5}{c}{\textbf{Gemma-4-E4B-it}} \\
\midrule
Base        & 22.91             & 54.3             & 21.2             & --             \\
LA          & 20.12             & 46.2             & 7.8              & --             \\
UM          & \textbf{18.94}    & \underline{67.1} & \underline{78.8} & --             \\
TPC         & 19.27             & \textbf{67.2}    & 78.6             & --             \\
ADS         & 19.75             & 66.0             & 75.9             & --             \\
TPC$\to$ADS & \underline{19.14} & 66.9             & \textbf{80.0}    & --             \\
\midrule
\rowcolor{gray!8}
$\Delta$ TPC vs.\ UM
  & \dloss{$+$0.33} & \dgain{$+$0.1} & \dloss{$-$0.2} & {--} \\
\rowcolor{gray!8}
$\Delta$ TPC$\to$ADS vs.\ TPC
  & \dgain{$-$0.13} & \dloss{$-$0.3} & \dgain{$+$1.4} & {--} \\
\midrule
\multicolumn{5}{c}{\textbf{Reference Models (no fine-tuning)}} \\
\midrule
Qwen2.5-Omni-3B  & 54.95 & 46.4 & 7.7  & 34.4 \\
GPT-audio        & 22.66 & 50.2 & 38.4 & 46.2 \\
Gemini-2.5-Pro   & 19.85 & 58.0 & 63.4 & 59.3 \\
\bottomrule
\end{tabular}
}
\vspace{-0.2cm}
\caption{
Results across generative and discriminative tasks by phases.
ASR reports WER ($\downarrow$); summarization: average BERTScore F1; DID and SER report
weighted-F1 ($\uparrow$). SER is evaluated on Qwen2.5-Omni-7B only.
The shaded $\Delta$ rows decompose scheduling effects in two steps:
$\Delta$ TPC vs.\ UM isolates the gain from curriculum ordering over the standard baseline,
and $\Delta$ TPC$\to$ADS vs.\ TPC isolates the additional effect of diversity-oriented sampling.
\dgain{Teal}: improvement; \dloss{orange}: degradation.
}
\label{tab:model-specific-phase-summary}
\vspace{-0.3cm}
\end{table}

\begin{figure}[!tbh]
\centering
\setlength{\tabcolsep}{0pt}
\begin{tabular}{c}
\includegraphics[width=0.45\textwidth,height=0.23\textheight,keepaspectratio]{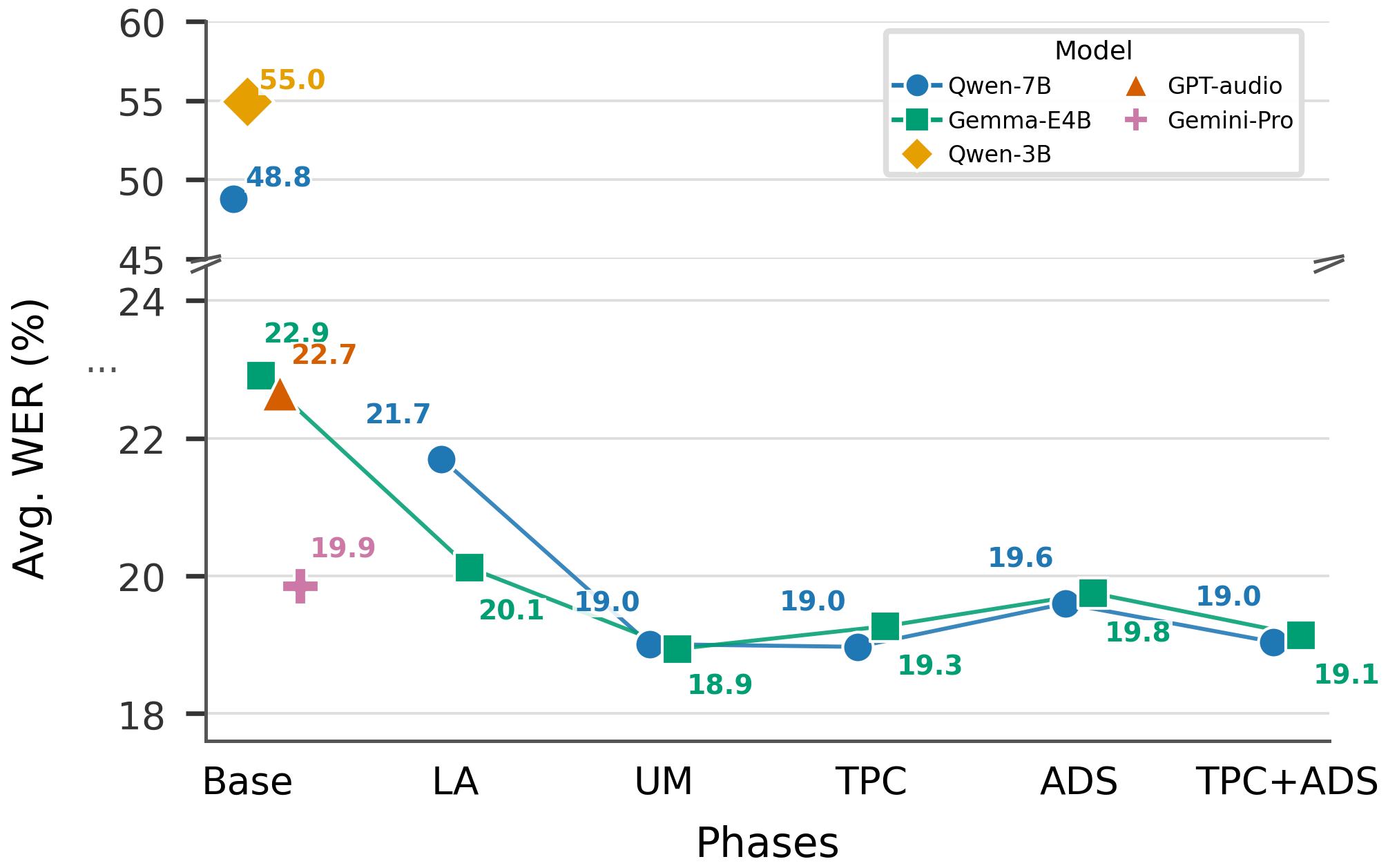} \\
\small \textbf{(a)} ASR WER \\
\vspace{-3mm} \\
\includegraphics[width=0.45\textwidth,height=0.23\textheight,keepaspectratio]{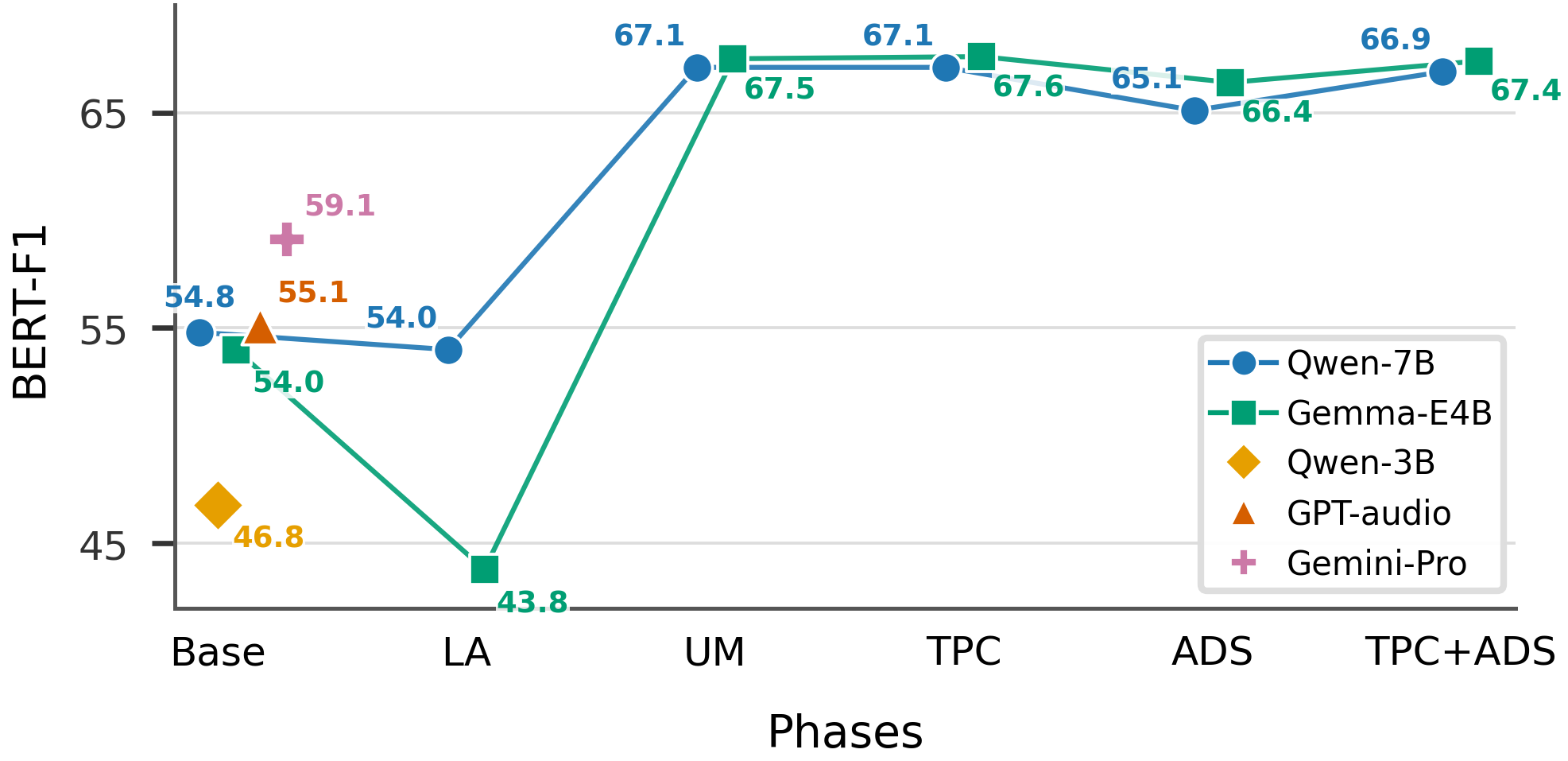} \\
\small \textbf{(b)} Summarization BERTScore F1 \\
\vspace{-3mm} \\
\includegraphics[width=0.5\textwidth,height=0.23\textheight,keepaspectratio]{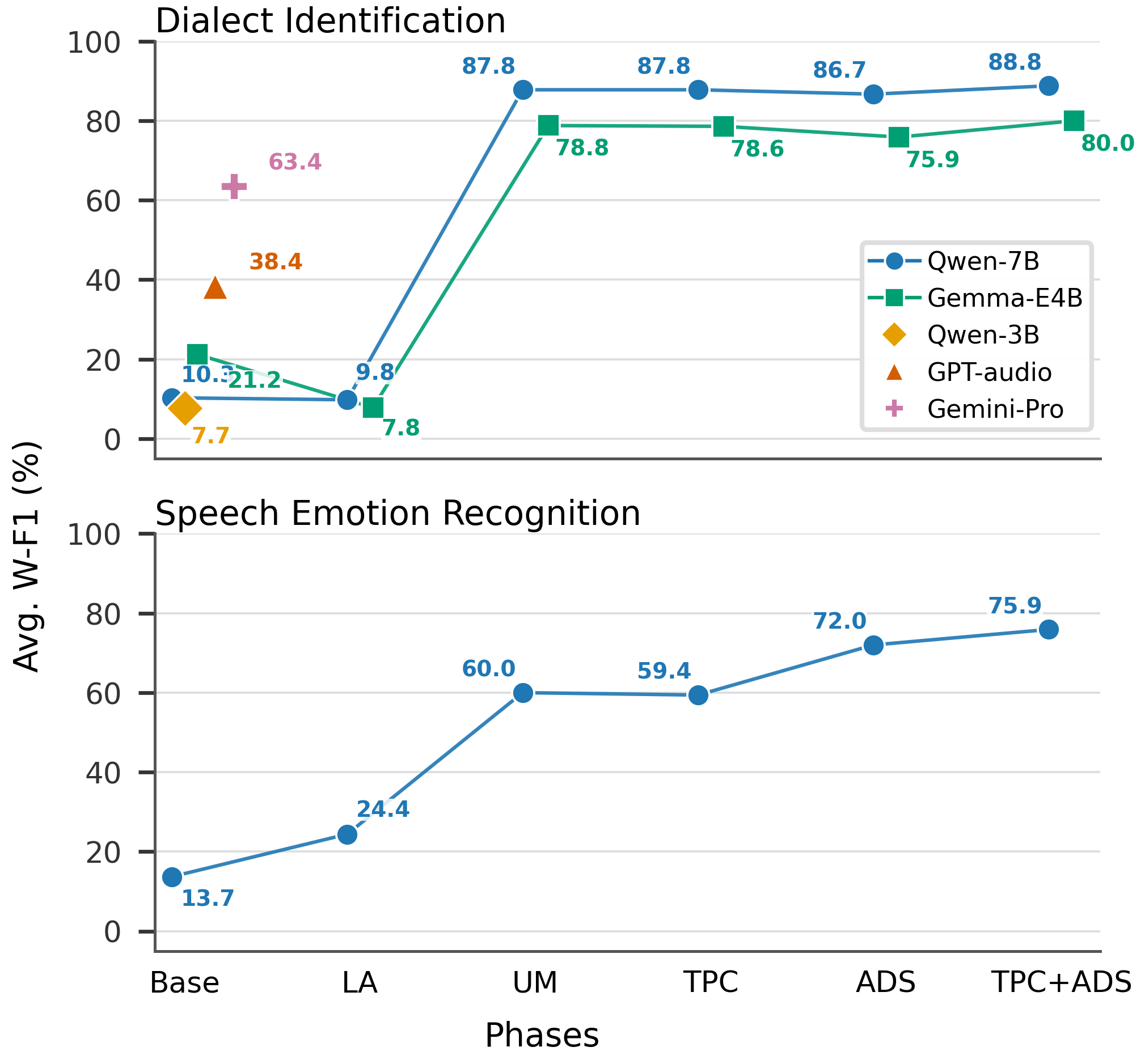} \\
\small \textbf{(c)} DID and SER macro-F1 \\
\end{tabular}
\vspace{-3mm}
\caption{
Phase-level results across task families. Panel (a) reports average ASR WER across MSA, dialectal Arabic, code-switched speech, native English, and L2 English. Panel (b) reports average summarization BERTScore F1 over SSUM and TSUM. Panel (c) reports DID and SER macro-F1. Lower is better for WER. Higher is better for all other metrics.
}
\label{fig-phase-results}
\vspace{-5mm}
\end{figure}

Figure~\ref{fig-phase-results} summarizes the same trends across task families. Panel (a) shows that adaptation sharply reduces ASR WER for Qwen2.5-Omni-7B. UM, TPC, and TPC$\rightarrow$ADS reach similar WER values near 19. Panel (b) shows that UM and TPC give the strongest summarization performance. ADS improves over the base models, but it remains below UM and TPC. Panel (c) shows a similar pattern for DID and SER. TPC$\rightarrow$ADS gives the strongest Qwen2.5-Omni-7B results on both tasks. These trends support a clear trade-off. UM and TPC provide stronger generative stability, while TPC$\rightarrow$ADS provides stronger discriminative robustness.

\subsection{Effect of Language-Centric Alignment}

LA gives a large ASR gain. It reduces Qwen2.5-Omni-7B WER from 48.79 to 21.70. It also reduces Gemma-4-E4B-it WER from 22.91 to 20.12. This shows that Phase 1 improves language-dependent acoustic and lexical alignment. However, LA alone does not solve higher-level speech understanding. DID and SER remain weak after LA, and summarization shows limited or negative gains. We therefore treat LA as a necessary alignment step, not as a complete adaptation strategy. Phase 2 multi-task instruction tuning is needed for semantic and paralinguistic behavior.

\subsection{Effect of Phase 2 Data Scheduling}
We compare UM, TPC, ADS, and TPC$\rightarrow$ADS under the same model, optimization settings, and training budget. This design isolates the effect of data scheduling and batch construction.

\noindent\textbf{Generative tasks.}
UM and TPC provide the strongest results for ASR and summarization. UM benefits from stable exposure to the natural task distribution. TPC starts with ASR and gives the model a strong audio-to-text mapping before adding higher-level tasks. This helps both ASR and summarization. ADS also improves over the base setting, but it usually trails UM and TPC on generative tasks. This suggests that ASR and summarization benefit from repeated exposure to common acoustic and lexical patterns.

\noindent\textbf{Discriminative tasks.}
DID and SER benefit more from label coverage and acoustic diversity. TPC$\rightarrow$ADS gives the best DID and SER results for Qwen2.5-Omni-7B. For Gemma-4-E4B-it, the TPC$\rightarrow$ADS gives the best DID score. We keep SER for Gemma as a future task.
These results show that diversity helps discriminative tasks, but its effect depends on the model and metric. They also show why ADS works better after TPC than as a standalone schedule. TPC first stabilizes the shared representation. ADS then improves coverage of labels, speakers, and acoustic conditions.

\noindent\textbf{Overall trade-off.}
Scheduling choices have a measurable impact under limited resources. UM and TPC
are preferable when ASR or summarization is the primary objective, as both strategies
provide stable generative performance with minimal tuning. TPC$\rightarrow$ADS is the
most reliable choice when discriminative robustness matters, combining early
stabilization with later diversity-aware sampling. Notably, TPC$\rightarrow$ADS
outperforms large proprietary reference models on discriminative tasks despite
operating at a fraction of their scale, surpassing Gemini-2.5-Pro on both DID (88.8
vs.\ 63.4) and SER (75.94 vs.\ 59.3), while remaining competitive on ASR and
summarization.


\section{Related Work}


\textbf{Audio and multimodal LLMs.} 
Recent work has extended LLMs to speech and other modalities by aligning pretrained encoders with language decoders. Early models such as SpeechT5 learned multiple speech tasks in a shared encoder--decoder framework \citep{ao2022speecht5}. Later systems, including AudioPaLM, SpeechGPT, and Qwen-Audio, combined LLM backbones with speech tokenization and instruction tuning for unified audio understanding and generation \citep{rubenstein2023audiopalm,zhang2023speechgpt,chu2023qwenaudio}. More recent omni models further scale this paradigm across modalities \citep{xu2025qwen25omnitechnicalreport,xu2025qwen3omnitechnicalreport,rouditchenko2025omni}. 
However, prior work gives limited attention to how data mixing affect stability and cross-task generalization in end-to-end audio instruction tuning.

\noindent
\textbf{Multi-task training.}
Multi-task audio instruction tuning is often affected by negative transfer, where dominant tasks suppress weaker or higher-level ones. Octavius addresses this problem with a LoRA-based mixture-of-experts design that reduces cross-task interference \citep{chen2024octavius}. MINT shows that naive task aggregation can harm generalization and that structured grouping improves transfer across tasks \citep{shan2025mint}. Curriculum learning has also been shown to improve optimization when the training order is well chosen \citep{bengio2009curriculum}. These findings motivate a systematic comparison between curriculum-style ordering and diversity-oriented sampling.


\noindent
\textbf{Arabic speech understanding.}
Arabic has benefited from multilingual speech foundation models such as Whisper and XLS-R, as well as LLM-based systems for Arabic ASR and dialect identification such as Octopus \citep{althubaiti-etal-2025-octopus}. However, higher-level Arabic speech understanding remains underdeveloped. Arabic summarization research is still largely text-based, for example AraBART \citep{eddine2022arabart}. Speech emotion recognition also continues to rely on relatively small task-specific datasets such as 
and 
KSUEmotions \citep{meftah2018ksuemotions}. Unified instruction-following audio models for Arabic remain largely unexplored.





\section{Conclusion}
We presented a controlled study of data scheduling and batch construction for 
multi-task instruction tuning of Arabic-centric omni audio LLMs, covering ASR, 
speech summarization, dialect identification, and emotion recognition. We also 
introduced \textbf{AraMega-SSum}, the first Arabic speech summarization benchmark. 
Our results confirm a clear efficiency--robustness trade-off: TPC provides the 
strongest generative performance, ADS improves paralinguistic robustness but hurts 
generative stability when used alone, and TPC$\rightarrow$ADS delivers the best 
overall balance across tasks. Notably, TPC$\rightarrow$ADS outperforms large 
proprietary models including Gemini-2.5-Pro on discriminative tasks, demonstrating 
that targeted scheduling can compensate for scale in low-resource, dialect-rich 
settings. Future work will extend this study across diverse AudioLLM architectures, 
compute budgets, and fine-tuning configurations.

\section*{Limitations}
Our study focuses on multi-task audio instruction tuning for a single AudioLLM, Qwen2.5-Omni 7B, in an Arabic--English setting. While this setup reflects a realistic and challenging deployment scenario, the findings may not fully generalize to models with substantially different architectures, scales, or training objectives. In particular, models with explicit task-specific heads or more decoupled audio--text pathways may respond differently to data scheduling strategies.

We evaluate scheduling strategies under a fixed training and a parameter-efficient fine-tuning setup. This choice is intentional and reflects practical compute constraints, but different conclusions may emerge under substantially larger budgets or full-parameter fine-tuning.

We compare four training strategies: uniform mixing, TPC, ADS, and TPC$\rightarrow$ADS. However, due to limited compute resources, we do not exhaustively evaluate all sampling combinations and ADS ablations, such as per-task uniform and label-balanced-only variants.

\section*{Societal/Broader Impact}
This work can improve Arabic speech technology in dialect-rich, code-switched settings, supporting accessibility (e.g., captions), education, and media/knowledge access with better transcription and spoken summarization. However, societal risks exist. Performance may vary across dialects, accents, and speaking styles, potentially creating unequal error rates and exclusion. Paralinguistic tasks (emotion/dialect recognition) can be misused if treated as definitive signals, and should not inform consequential decisions without strong validation and human oversight. Finally, AraMega-SSum relies on translation and TTS/voice cloning, which raises source-traceability and misuse concerns (e.g., impersonation) if not clearly documented. We therefore emphasize responsible release and use: transparent dataset documentation (licensing, consent, and “synthetic” labeling), subgroup evaluation, and explicit restrictions against surveillance or high-stakes deployment without safeguards.

\section*{Ethical Considerations}
Our work leverages large-scale Arabic and English speech corpora, including synthetic audio generated via neural voice cloning. While these methods enable the creation of diverse and balanced datasets, they raise important ethical considerations. First, the use of voice cloning and synthesized speech could potentially be misused for impersonation, fraud, or other malicious purposes. Second, even when using publicly available or synthetic data, issues of privacy and consent must be carefully managed, particularly when real speaker recordings are involved. We mitigate these concerns by curating datasets with explicit permissions where applicable, ensuring anonymization of speaker identities, and restricting dataset release to research purposes only. Future work should continue to address these risks, especially in low-resource and dialect-rich contexts where consent practices may vary.

\bibliography{custom}

@article{alam2025everydaymmqa,
  title = {{OASIS}: A Multilingual and Multimodal Framework for Culturally Grounded Spoken Visual QA},
  author = {Alam, Firoj and Shahroor, Ali Ezzat and Hasan, Md. Arid and Ali, Zien Sheikh and Bhatti, Hunzalah Hassan and Kmainasi, Mohamed Bayan and Chowdhury, Shammur Absar and Mousi, Basel and Dalvi, Fahim and Durrani, Nadir and Milic-Frayling, Natasa},
  journal = {arXiv preprint arXiv:2510.06371},
  year = {2025},
}

@inproceedings{radford2023robust,
  title={Robust speech recognition via large-scale weak supervision},
  author={Radford, Alec and Kim, Jong Wook and Xu, Tao and Brockman, Greg and McLeavey, Christine and Sutskever, Ilya},
  booktitle={International conference on machine learning},
  pages={28492--28518},
  year={2023},
  organization={PMLR}
}

@article{du2025cosyvoice,
  title={{CosyVoice 3}: Towards in-the-wild speech generation via scaling-up and post-training},
  author={Du, Zhihao and Gao, Changfeng and Wang, Yuxuan and Yu, Fan and Zhao, Tianyu and Wang, Hao and Lv, Xiang and Wang, Hui and Ni, Chongjia and Shi, Xian and others},
  journal={arXiv preprint arXiv:2505.17589},
  year={2025}
}

@inproceedings{10.5555/3692070.3692979,
author = {Ju, Zeqian and Wang, Yuancheng and Shen, Kai and Tan, Xu and Xin, Detai and Yang, Dongchao and Liu, Yanqing and Leng, Yichong and Song, Kaitao and Tang, Siliang and Wu, Zhizheng and Qin, Tao and Li, Xiang-Yang and Ye, Wei and Zhang, Shikun and Bian, Jiang and He, Lei and Li, Jinyu and Zhao, Sheng},
title = {NaturalSpeech 3: zero-shot speech synthesis with factorized codec and diffusion models},
year = {2024},
publisher = {JMLR.org},
booktitle = {Proceedings of the 41st International Conference on Machine Learning},
articleno = {909},
numpages = {19},
location = {Vienna, Austria},
series = {ICML'24}
}

@article{gwet2008computing,
 author = {Gwet, Kilem Li},
 journal = {British Journal of Mathematical and Statistical Psychology},
 number = {1},
 pages = {29--48},
 publisher = {Wiley Online Library},
 title = {Computing inter-rater reliability and its variance in the presence of high agreement},
 volume = {61},
 year = {2008}
}

@article{ali2026menaspeechbankreferencevoicebank,
  title = {{MENASpeechBank}: A Reference Voice Bank with Persona-Conditioned Multi-Turn Conversations for AudioLLMs},
  author = {Ali, Zien Sheikh and Bhatti, Hunzalah Hassan and Nandi, Rabindra Nath and Chowdhury, Shammur Absar and Alam, Firoj},
  journal = {arXiv preprint arXiv:2602.07036},
  year = {2026},
  url = {https://arxiv.org/abs/2602.07036},
}

@article{zheng2023judging, 
  title={Judging {LLM}-as-a-judge with {MT-Bench} and {Chatbot Arena}},
  author={Zheng, Lianmin and Chiang, Wei-Lin and Sheng, Ying and Zhuang, Siyuan and Wu, Zhanghao and Zhuang, Yonghao and Lin, Zi and Li, Zhuohan and Li, Dacheng and Xing, Eric and others},
  journal={Advances in Neural Information Processing Systems},
  volume={36},
  pages={46595--46623},
  year={2023}
}

@inproceedings{lin2004rouge,
    title = "{ROUGE}: A Package for Automatic Evaluation of Summaries",
    author = "Lin, Chin-Yew",
    booktitle = "Text Summarization Branches Out",
    month = jul,
    year = "2004",
    address = "Barcelona, Spain",
    publisher = "Association for Computational Linguistics",
    url = "https://aclanthology.org/W04-1013/",
    pages = "74--81"
}

@dataset{klaylat2018arabic,
  author       = {Klaylat, Samira and Osman, Ziad and Zantout, Rached and Hamandi, Lama},
  title        = {Arabic Natural Audio Dataset},
  year         = {2018},
  publisher    = {Mendeley Data},
  version      = {V1},
  doi          = {10.17632/xm232yxf7t.1},
  url          = {https://doi.org/10.17632/xm232yxf7t.1}
}

@INPROCEEDINGS{kedas2022,
  author={Belhadj, Mourad and Bendellali, Ilham and Lakhdari, Elalia},
  booktitle={2022 International Symposium on iNnovative Informatics of Biskra (ISNIB)}, 
  title={{KEDAS}: A validated Arabic Speech Emotion Dataset}, 
  year={2022},
  volume={},
  number={},
  pages={1-6},
  doi={10.1109/ISNIB57382.2022.10075694}
}

@data{eaed2024,
doi = {10.21227/3w6s-ft91},
url = {https://dx.doi.org/10.21227/3w6s-ft91},
author = {Sarah Safwat and Mohammed Salem and Nada Sharaf},
publisher = {IEEE Dataport},
title = {Egyptian Arabic Emotional Dataset ({EAED})},
year = {2023} 
}

@inproceedings{ao2022speecht5,
    title = "{S}peech{T}5: Unified-Modal Encoder-Decoder Pre-Training for Spoken Language Processing",
    author = "Ao, Junyi  and
      Wang, Rui  and
      Zhou, Long  and
      Wang, Chengyi  and
      Ren, Shuo  and
      Wu, Yu  and
      Liu, Shujie  and
      Ko, Tom  and
      Li, Qing  and
      Zhang, Yu  and
      Wei, Zhihua  and
      Qian, Yao  and
      Li, Jinyu  and
      Wei, Furu",
    editor = "Muresan, Smaranda  and
      Nakov, Preslav  and
      Villavicencio, Aline",
    booktitle = "Proceedings of the 60th Annual Meeting of the Association for Computational Linguistics (Volume 1: Long Papers)",
    month = may,
    year = "2022",
    address = "Dublin, Ireland",
    publisher = "Association for Computational Linguistics",
    url = "https://aclanthology.org/2022.acl-long.393/",
    doi = "10.18653/v1/2022.acl-long.393",
    pages = "5723--5738"
}

@inproceedings{zhang2023speechgpt,
    title = "{S}peech{GPT}: Empowering Large Language Models with Intrinsic Cross-Modal Conversational Abilities",
    author = "Zhang, Dong  and
      Li, Shimin  and
      Zhang, Xin  and
      Zhan, Jun  and
      Wang, Pengyu  and
      Zhou, Yaqian  and
      Qiu, Xipeng",
    editor = "Bouamor, Houda  and
      Pino, Juan  and
      Bali, Kalika",
    booktitle = "Findings of the Association for Computational Linguistics: EMNLP 2023",
    month = dec,
    year = "2023",
    address = "Singapore",
    publisher = "Association for Computational Linguistics",
    url = "https://aclanthology.org/2023.findings-emnlp.1055/",
    doi = "10.18653/v1/2023.findings-emnlp.1055",
    pages = "15757--15773"
}

@article{chu2023qwenaudio,
  title={{Qwen-Audio}: Advancing Universal Audio Understanding via Unified Large-Scale Audio-Language Models},
  author={Chu, Yunfei and Xu, Jin and Zhou, Xiaohuan and Yang, Qian and Zhang, Shiliang and Yan, Zhijie and Zhou, Chang and Zhou, Jingren},
  journal={arXiv preprint arXiv:2311.07919},
  year={2023},
  eprint={2311.07919},
  archivePrefix={arXiv},
  primaryClass={eess.AS},
  url={https://arxiv.org/abs/2311.07919}
}

@article{shan2025mint,
  title={{MINT}: Multimodal Instruction Tuning with Multimodal Interaction Grouping},
  author={Shan, Xiaojun and Cao, Qi and Han, Xing and Yu, Haofei and Liang, Paul Pu},
  journal={arXiv preprint arXiv:2506.02308},
  year={2025},
  eprint={2506.02308},
  archivePrefix={arXiv},
  primaryClass={cs.LG},
  url={https://arxiv.org/abs/2506.02308}
}

@inproceedings{eddine2022arabart,
    title = "{A}ra{BART}: a Pretrained {A}rabic Sequence-to-Sequence Model for Abstractive Summarization",
    author = "Kamal Eddine, Moussa  and
      Tomeh, Nadi  and
      Habash, Nizar  and
      Le Roux, Joseph  and
      Vazirgiannis, Michalis",
    editor = "Bouamor, Houda  and
      Al-Khalifa, Hend  and
      Darwish, Kareem  and
      Rambow, Owen  and
      Bougares, Fethi  and
      Abdelali, Ahmed  and
      Tomeh, Nadi  and
      Khalifa, Salam  and
      Zaghouani, Wajdi",
    booktitle = "Proceedings of the Seventh Arabic Natural Language Processing Workshop (WANLP)",
    month = dec,
    year = "2022",
    address = "Abu Dhabi, United Arab Emirates (Hybrid)",
    publisher = "Association for Computational Linguistics",
    url = "https://aclanthology.org/2022.wanlp-1.4/",
    doi = "10.18653/v1/2022.wanlp-1.4",
    pages = "31--42"
}

@inproceedings{chen2024octavius,
  title={Octavius: Mitigating task interference in mllms via lora-moe},
  author={Chen, Zeren and Liu, Huayang and Yin, Zhenfei and Liu, Si and Sheng, Lyu and Ouyang, Wanli and Shao, Jing and others},
  booktitle={International Conference on Learning Representations},
  volume={2024},
  pages={24652--24673},
  year={2024}
}

@inproceedings{casanova2024xtts,
  title     = {{XTTS: a Massively Multilingual Zero-Shot Text-to-Speech Model}},
  author    = {Edresson Casanova and Kelly Davis and Eren Gölge and Görkem Göknar and Iulian Gulea and Logan Hart and Aya Aljafari and Joshua Meyer and Reuben Morais and Samuel Olayemi and Julian Weber},
  year      = {2024},
  booktitle = {{Interspeech 2024}},
  pages     = {4978--4982},
  doi       = {10.21437/Interspeech.2024-2016},
  issn      = {2958-1796},
}

@inproceedings{matsuura24_interspeech,
  title     = {{Sentence-wise Speech Summarization: Task, Datasets, and End-to-End Modeling with LM Knowledge Distillation}},
  author    = {Kohei Matsuura and Takanori Ashihara and Takafumi Moriya and Masato Mimura and Takatomo Kano and Atsunori Ogawa and Marc Delcroix},
  year      = {2024},
  booktitle = {{Interspeech 2024}},
  pages     = {1945--1949},
  doi       = {10.21437/Interspeech.2024-349},
  issn      = {2958-1796},
}

@INPROCEEDINGS{ali2016mgb2,
  author={Ali, Ahmed and Bell, Peter and Glass, James and Messaoui, Yacine and Mubarak, Hamdy and Renals, Steve and Zhang, Yifan},
  booktitle={2016 IEEE Spoken Language Technology Workshop (SLT)}, 
  title={The {MGB-2} challenge: Arabic multi-dialect broadcast media recognition}, 
  year={2016},
  volume={},
  number={},
  pages={279-284},
  doi={10.1109/SLT.2016.7846277}
  }

@ARTICLE{alotaibi2020ksu,
  author={Meftah, Ali Hamid and Qamhan, Mustafa A. and Seddiq, Yasser and Alotaibi, Yousef A. and Selouani, Sid Ahmed},
  journal={IEEE Access}, 
  title={King Saud University Emotions Corpus: Construction, Analysis, Evaluation, and Comparison}, 
  year={2021},
  volume={9},
  number={},
  pages={54201-54219},
  doi={10.1109/ACCESS.2021.3070751}}

@article{xu2025qwen25omnitechnicalreport,
  title={Qwen2.5-Omni Technical Report},
  author={Xu, Jin and Guo, Zhifang and He, Jinzheng and Hu, Hangrui and He, Ting and Bai, Shuai and Chen, Keqin and Wang, Jialin and Fan, Yang and Dang, Kai and Zhang, Bin and Wang, Xiong and Chu, Yunfei and Lin, Junyang},
  journal={arXiv preprint arXiv:2503.20215},
  year={2025},
  url={https://arxiv.org/abs/2503.20215}
}

@inproceedings{musan2015,
  title={{MUSAN}: {A} {M}ultipurpose {C}orpus for {M}usic and {N}oise},
  author={Snyder, David and Chen, Guoguo and Povey, Daniel},
  booktitle={arXiv preprint arXiv:1510.08484},
  year={2015}
}

@inproceedings{ko2017rir,
author = {Ko, Tom and Peddinti, Vijayaditya and Seltzer, Michael and Khudanpur, Sanjeev},
year = {2017},
month = {03},
pages = {5220-5224},
title = {A study on data augmentation of reverberant speech for robust speech recognition},
doi = {10.1109/ICASSP.2017.7953152}
}

@INPROCEEDINGS{panayotov2015librispeech,
  author={Panayotov, Vassil and Chen, Guoguo and Povey, Daniel and Khudanpur, Sanjeev},
  booktitle={2015 IEEE International Conference on Acoustics, Speech and Signal Processing (ICASSP)}, 
  title={Librispeech: An ASR corpus based on public domain audio books}, 
  year={2015},
  volume={},
  number={},
  pages={5206-5210},
  doi={10.1109/ICASSP.2015.7178964}
  }

@INPROCEEDINGS{alharbi2024sada,
  author={Alharbi, Sadeen and Alowisheq, Areeb and Tüske, Zoltán and Darwish, Kareem and Alrajeh, Abdullah and Alrowithi, Abdulmajeed and Tamran, Aljawharah Bin and Ibrahim, Asma and Aloraini, Raghad and Alnajim, Raneem and Alkahtani, Ranya and Almuasaad, Renad and Alrasheed, Sara and Alsubaie, Shaykhah and Alonaizan, Yaser},
  booktitle={ICASSP 2024 - 2024 IEEE International Conference on Acoustics, Speech and Signal Processing (ICASSP)}, 
  title={{SADA}: Saudi Audio Dataset for Arabic}, 
  year={2024},
  volume={},
  number={},
  pages={10286-10290},
  doi={10.1109/ICASSP48485.2024.10446243}
  }

@inproceedings{chowdhury2020dacs,
  title     = {{Effects of Dialectal Code-Switching on Speech Modules: A Study Using Egyptian Arabic Broadcast Speech}},
  author    = {Shammur A. Chowdhury and Younes Samih and Mohamed Eldesouki and Ahmed Ali},
  year      = {2020},
  booktitle = {{Interspeech 2020}},
  pages     = {2382--2386},
  doi       = {10.21437/Interspeech.2020-2271},
  issn      = {2958-1796},
}

@inproceedings{zhang2019bertscore,
title={{BERTScore}: Evaluating Text Generation with BERT},
author={Tianyi Zhang and Varsha Kishore and Felix Wu and Kilian Q. Weinberger and Yoav Artzi},
booktitle={International Conference on Learning Representations},
year={2020},
url={https://openreview.net/forum?id=SkeHuCVFDr}
}

@INPROCEEDINGS{ali2017mgb3,
  author={Ali, Ahmed and Vogel, Stephan and Renals, Steve},
  booktitle={2017 IEEE Automatic Speech Recognition and Understanding Workshop (ASRU)}, 
  title={Speech recognition challenge in the wild: Arabic MGB-3}, 
  year={2017},
  volume={},
  number={},
  pages={316-322},
  doi={10.1109/ASRU.2017.8268952}}

@inproceedings{ali2021mgb5,
  title={The {MGB} challenge: Recognition and dialect identification of dialectal arabic speech},
  author={Ali, Ahmed and Shon, Suwon and Samih, Younes and Mubarak, Hamdy and Abdelali, Ahmed and Glass, James and Renals, Steve and Choukri, Khalid},
  booktitle={2019 IEEE Automatic Speech Recognition and Understanding Workshop (ASRU)},
  pages={1026--1033},
  year={2019},
  organization={IEEE}
}

@INPROCEEDINGS{fan2020adi17,
  author={Shon, Suwon and Ali, Ahmed and Samih, Younes and Mubarak, Hamdy and Glass, James},
  booktitle={ICASSP 2020 - 2020 IEEE International Conference on Acoustics, Speech and Signal Processing (ICASSP)}, 
  title={ADI17: A Fine-Grained Arabic Dialect Identification Dataset}, 
  year={2020},
  volume={},
  number={},
  pages={8244-8248},
  doi={10.1109/ICASSP40776.2020.9052982}
  }

@INPROCEEDINGS{al-fetyani-etal-2023-masc,
  author={Al-Fetyani, Mohammad and Al-Barham, Muhammad and Abandah, Gheith and Alsharkawi, Adham and Dawas, Maha},
  booktitle={2022 IEEE Spoken Language Technology Workshop (SLT)}, 
  title={{MASC}: Massive Arabic Speech Corpus}, 
  year={2023},
  volume={},
  number={},
  pages={1006-1013},
  doi={10.1109/SLT54892.2023.10022652}}

@inproceedings{mubarak2021qasr,
    title = "{QASR}: {QCRI} Aljazeera Speech Resource A Large Scale Annotated {A}rabic Speech Corpus",
    author = "Mubarak, Hamdy  and
      Hussein, Amir  and
      Chowdhury, Shammur Absar  and
      Ali, Ahmed",
    editor = "Zong, Chengqing  and
      Xia, Fei  and
      Li, Wenjie  and
      Navigli, Roberto",
    booktitle = "Proceedings of the 59th Annual Meeting of the Association for Computational Linguistics and the 11th International Joint Conference on Natural Language Processing (Volume 1: Long Papers)",
    month = aug,
    year = "2021",
    address = "Online",
    publisher = "Association for Computational Linguistics",
    url = "https://aclanthology.org/2021.acl-long.177/",
    doi = "10.18653/v1/2021.acl-long.177",
    pages = "2274--2285"
}

@inproceedings{chen2021gigaspeech,
   title={{GigaSpeech}: An Evolving, Multi-Domain ASR Corpus with 10,000 Hours of Transcribed Audio},
   url={http://dx.doi.org/10.21437/Interspeech.2021-1965},
   DOI={10.21437/interspeech.2021-1965},
   booktitle={Interspeech 2021},
   publisher={ISCA},
   author={Chen, Guoguo and Chai, Shuzhou and Wang, Guan-Bo and Du, Jiayu and Zhang, Wei-Qiang and Weng, Chao and Su, Dan and Povey, Daniel and Trmal, Jan and Zhang, Junbo and Jin, Mingjie and Khudanpur, Sanjeev and Watanabe, Shinji and Zhao, Shuaijiang and Zou, Wei and Li, Xiangang and Yao, Xuchen and Wang, Yongqing and You, Zhao and Yan, Zhiyong},
   year={2021},
   month=aug, collection={interspeech_2021} 
}

@inproceedings{ardila-etal-2020-common,
  title={{Common Voice}: A massively-multilingual speech corpus},
  author={Ardila, Rosana and Branson, Megan and Davis, Kelly and Kohler, Michael and Meyer, Josh and Henretty, Michael and Morais, Reuben and Saunders, Lindsay and Tyers, Francis and Weber, Gregor},
  booktitle={Proceedings of the twelfth language resources and evaluation conference},
  pages={4218--4222},
  year={2020}
}

@inproceedings{zhao2018l2arctic,
  title     = {{L2-ARCTIC: A Non-native English Speech Corpus}},
  author    = {Guanlong Zhao and Sinem Sonsaat and Alif Silpachai and Ivana Lucic and Evgeny Chukharev-Hudilainen and John Levis and Ricardo Gutierrez-Osuna},
  year      = {2018},
  booktitle = {{Interspeech 2018}},
  pages     = {2783--2787},
  doi       = {10.21437/Interspeech.2018-1110},
  issn      = {2958-1796},
}

@inproceedings{o2021spgispeech,
  title     = {{SPGISpeech: 5,000 Hours of Transcribed Financial Audio for Fully Formatted End-to-End Speech Recognition}},
  author    = {Patrick K. O’Neill and Vitaly Lavrukhin and Somshubra Majumdar and Vahid Noroozi and Yuekai Zhang and Oleksii Kuchaiev and Jagadeesh Balam and Yuliya Dovzhenko and Keenan Freyberg and Michael D. Shulman and Boris Ginsburg and Shinji Watanabe and Georg Kucsko},
  year      = {2021},
  booktitle = {{Interspeech 2021}},
  pages     = {1434--1438},
  doi       = {10.21437/Interspeech.2021-1860},
  issn      = {2958-1796},
}

@inbook{hernandez2018ted,
   title={{TED-LIUM 3}: Twice as Much Data and Corpus Repartition for Experiments on Speaker Adaptation},
   ISBN={9783319995793},
   ISSN={1611-3349},
   url={http://dx.doi.org/10.1007/978-3-319-99579-3_21},
   DOI={10.1007/978-3-319-99579-3_21},
   booktitle={Speech and Computer},
   publisher={Springer International Publishing},
   author={Hernandez, François and Nguyen, Vincent and Ghannay, Sahar and Tomashenko, Natalia and Estève, Yannick},
   year={2018},
   pages={198–208} }

@inproceedings{chowdhury_towards_2021,
	title = {Towards {One} {Model} to {Rule} {All}: {Multilingual} {Strategy} for {Dialectal} {Code}-{Switching} {Arabic} {ASR}},
	booktitle = {{{{Proc. of INTERSPEECH}}}},
	author = {Chowdhury, Shammur Absar and Hussein, Amir and Abdelali, Ahmed and Ali, Ahmed},
	year = {2021}
}

@article{rouditchenko2025omni,
  title={{Omni-R1}: Do You Really Need Audio to Fine-Tune Your Audio LLM?},
  author={Rouditchenko, Andrew and Bhati, Saurabhchand and Araujo, Edson and Thomas, Samuel and Kuehne, Hilde and Feris, Rogerio and Glass, James},
  journal={arXiv preprint arXiv:2505.09439},
  year={2025}
}

@article{xu2025qwen3omnitechnicalreport,
  title={Qwen3-Omni Technical Report},
  author={Xu, Jin and Guo, Zhifang and Hu, Hangrui and Chu, Yunfei and Wang, Xiong and He, Jinzheng and Wang, Yuxuan and Shi, Xian and He, Ting and Zhu, Xinfa and Lv, Yuanjun and Wang, Yongqi and Guo, Dake and Wang, He and Ma, Linhan and Zhang, Pei and Zhang, Xinyu and Hao, Hongkun and Guo, Zishan and Yang, Baosong and Zhang, Bin and Ma, Ziyang and Wei, Xipin and Bai, Shuai and Chen, Keqin and Liu, Xuejing and Wang, Peng and Yang, Mingkun and Liu, Dayiheng and Ren, Xingzhang and Zheng, Bo and Men, Rui and Zhou, Fan and Yu, Bowen and Yang, Jianxin and Yu, Le and Zhou, Jingren and Lin, Junyang},
  journal={arXiv preprint arXiv:2509.17765},
  year={2025},
  eprint={2509.17765},
  archivePrefix={arXiv},
  primaryClass={cs.CL},
  url={https://arxiv.org/abs/2509.17765}
}

@article{rubenstein2023audiopalm,
  title={{AudioPaLM}: A Large Language Model That Can Speak and Listen},
  author={Rubenstein, Paul K. and Asawaroengchai, Chulayuth and Nguyen, Duc Dung and Bapna, Ankur and Borsos, Zal{\'a}n and de Chaumont Quitry, F{\'e}lix and Chen, Peter and El Badawy, Dalia and Han, Wei and Kharitonov, Eugene and Muckenhirn, Hannah and Padfield, Dirk and Qin, James and Rozenberg, Danny and Sainath, Tara and Schalkwyk, Johan and Sharifi, Matt and Ramanovich, Michelle Tadmor and Tagliasacchi, Marco and Tudor, Alexandru and Velimirovi{\'c}, Mihajlo and Vincent, Damien and Yu, Jiahui and Wang, Yongqiang and Zayats, Vicky and Zeghidour, Neil and Zhang, Yu and Zhang, Zhishuai and Zilka, Lukas and Frank, Christian},
  journal={arXiv preprint arXiv:2306.12925},
  year={2023}
}

@inproceedings{bengio2009curriculum,
  title        = {Curriculum Learning},
  author       = {Bengio, Yoshua and Louradour, J{\'e}r{\^o}me and Collobert, Ronan and Weston, Jason},
  booktitle    = {Proceedings of the 26th International Conference on Machine Learning (ICML)},
  year         = {2009},
  pages        = {41--48},
  publisher    = {ACM},
  doi          = {10.1145/1553374.1553380}
}

@article{meftah2018ksuemotions,
  title={Evaluation of an Arabic speech corpus of emotions: A perceptual and statistical analysis},
  author={Meftah, Ali Hamid and Alotaibi, Yousef Ajami and Selouani, Sid-Ahmed},
  journal={IEEE Access},
  volume={6},
  pages={72845--72861},
  year={2018},
  publisher={IEEE}
}

@inproceedings{althubaiti-etal-2025-octopus,
    title = "Octopus: Towards Building the {A}rabic Speech {LLM} Suite",
    author = "Althubaiti, Sara  and
      Lodagala, Vasista Sai  and
      Clark, Tjad  and
      Elshahawy, Yousseif Ahmed  and
      Izham, Daniel  and
      Alrajeh, Abdullah  and
      Bin Tamran, Aljawahrah  and
      Ali, Ahmed",
    editor = "Darwish, Kareem  and
      Ali, Ahmed  and
      Abu Farha, Ibrahim  and
      Touileb, Samia  and
      Zitouni, Imed  and
      Abdelali, Ahmed  and
      Al-Ghamdi, Sharefah  and
      Alkhereyf, Sakhar  and
      Zaghouani, Wajdi  and
      Khalifa, Salam  and
      AlKhamissi, Badr  and
      Almatham, Rawan  and
      Hamed, Injy  and
      Alyafeai, Zaid  and
      Alowisheq, Areeb  and
      Inoue, Go  and
      Mrini, Khalil  and
      Alshammari, Waad",
    booktitle = "Proceedings of The Third Arabic Natural Language Processing Conference",
    month = nov,
    year = "2025",
    address = "Suzhou, China",
    publisher = "Association for Computational Linguistics",
    url = "https://aclanthology.org/2025.arabicnlp-main.35/",
    doi = "10.18653/v1/2025.arabicnlp-main.35",
    pages = "425--435",
    ISBN = "979-8-89176-352-4"
}

@article{zhang2023google,
  title={Google usm: Scaling automatic speech recognition beyond 100 languages},
  author={Zhang, Yu and Han, Wei and Qin, James and Wang, Yongqiang and Bapna, Ankur and Chen, Zhehuai and Chen, Nanxin and Li, Bo and Axelrod, Vera and Wang, Gary and others},
  journal={arXiv preprint arXiv:2303.01037},
  year={2023}
}

\appendix
\section*{Appendix}

\section{Dataset Details}
\label{sec_app_dataset_details}

\begin{table}[t]
\centering
\scriptsize
\setlength{\tabcolsep}{3pt}
\renewcommand{\arraystretch}{0.94}
\begin{tabularx}{\columnwidth}{@{} l l X r @{}}
\toprule
\textbf{Obj.} & \textbf{Task} & \textbf{Data} & \textbf{Total} \\
\midrule
\rowcolor{phasecolor}
\multicolumn{4}{c}{\textbf{Phase 1: Language-centric Alignment}} \\
\midrule
ASR & ASR &
QASR~\cite{mubarak2021qasr}, MASC~\cite{al-fetyani-etal-2023-masc}, MGB-3/5~\cite{ali2017mgb3,ali2021mgb5}, GigaSpeech~\cite{chen2021gigaspeech}, GALE,\footnotemark[1] TED-LIUM~3~\cite{hernandez2018ted}, LibriSpeech~\cite{panayotov2015librispeech}, CV (ar/en)~\cite{ardila-etal-2020-common}, SPGISpeech~\cite{o2021spgispeech}, + in-house
& 12,000h \\
ASR & ASR & Augmentation & $\approx$15,000h \\
\rowcolor{totalcolor}
\textbf{Total} & & & $\approx$\textbf{27,000h} \\
\midrule
\rowcolor{phasecolor}
\multicolumn{4}{c}{\textbf{Phase 2: Multi-task Training}} \\
\midrule
Gen. & ASR & $\sim$15\% subset of Phase~1 & $\approx$1,740h \\
Gen. & SSUM & MegaSSUM~\cite{matsuura24_interspeech} + AraMega-SSum (ours) & 374h \\
Gen. & TSUM & MegaSSUM~\cite{matsuura24_interspeech} + AraMega-SSum (ours) & 99,995\# \\
Disc. & DID & ADI-17~\cite{fan2020adi17} + ADI-5 (MSA only)~\cite{ali2017mgb3} & $\approx$2,983h \\
Disc. & Emo. & ANAD~\cite{klaylat2018arabic} + EAED~\cite{eaed2024} + KEDAS~\cite{kedas2022} + KSU~\cite{alotaibi2020ksu} 
& 12.17h \\
\rowcolor{totalcolor}
\textbf{Total} & & & \textbf{5,109h + 99,995\#} \\
\bottomrule
\end{tabularx}
\vspace{-0.3cm}
\caption{Datasets used in each training phase. Obj.denotes task objective, Gen.: generative, Disc.: discriminative, SSUM: speech summarization, TSUM: text summarization, DID: dialect identification, Emo.: emotion recognition, CV: Common Voice. ``h'' denotes audio duration and ``\#'' denotes the number of samples.}
\label{tab:dataset_stats}
\vspace{-0.3cm}
\end{table}

\begin{table}[t]
\centering
\scriptsize
\setlength{\tabcolsep}{4pt}
\renewcommand{\arraystretch}{0.94}
\begin{tabularx}{\columnwidth}{@{} l X l @{}}
\toprule
\textbf{Task} & \textbf{Benchmarks} & \textbf{Metric} \\
\midrule
ASR (Nat.) & LibriSpeech~\cite{panayotov2015librispeech}, MGB-2~\cite{ali2016mgb2} & WER \\
ASR (Dial.) & MGB-3~\cite{ali2017mgb3}, SADA~\cite{alharbi2024sada} & WER \\
ASR (L2/CS) & L2-ARCTIC~\cite{zhao2018l2arctic}, ESCWA~\cite{chowdhury_towards_2021}, DACS~\cite{chowdhury2020dacs} & WER \\
Summ. & MegaSUM-SSum~\cite{matsuura24_interspeech}, AraMega-SSum (\textbf{ours}) & RL, BSc., Lj \\
Emotion & KSU~\cite{alotaibi2020ksu} 
& w-F1 \\
DID & ADI-17~\cite{fan2020adi17} & w-F1 \\
\bottomrule
\end{tabularx}
\vspace{-0.3cm}
\caption{Evaluation datasets and metrics. Nat. denotes native, Dial.: dialectal, L2: second-language, and CS: code-switching. Summ.: summarization, RL: ROUGE-L, BSc.: BERTScore, Lj: LLM-judge and w-F1: weighted F1.}
\label{tab:benchmarks}
\vspace{-0.3cm}
\end{table}

\begin{figure}
    \centering
    \includegraphics[width=0.95\linewidth]{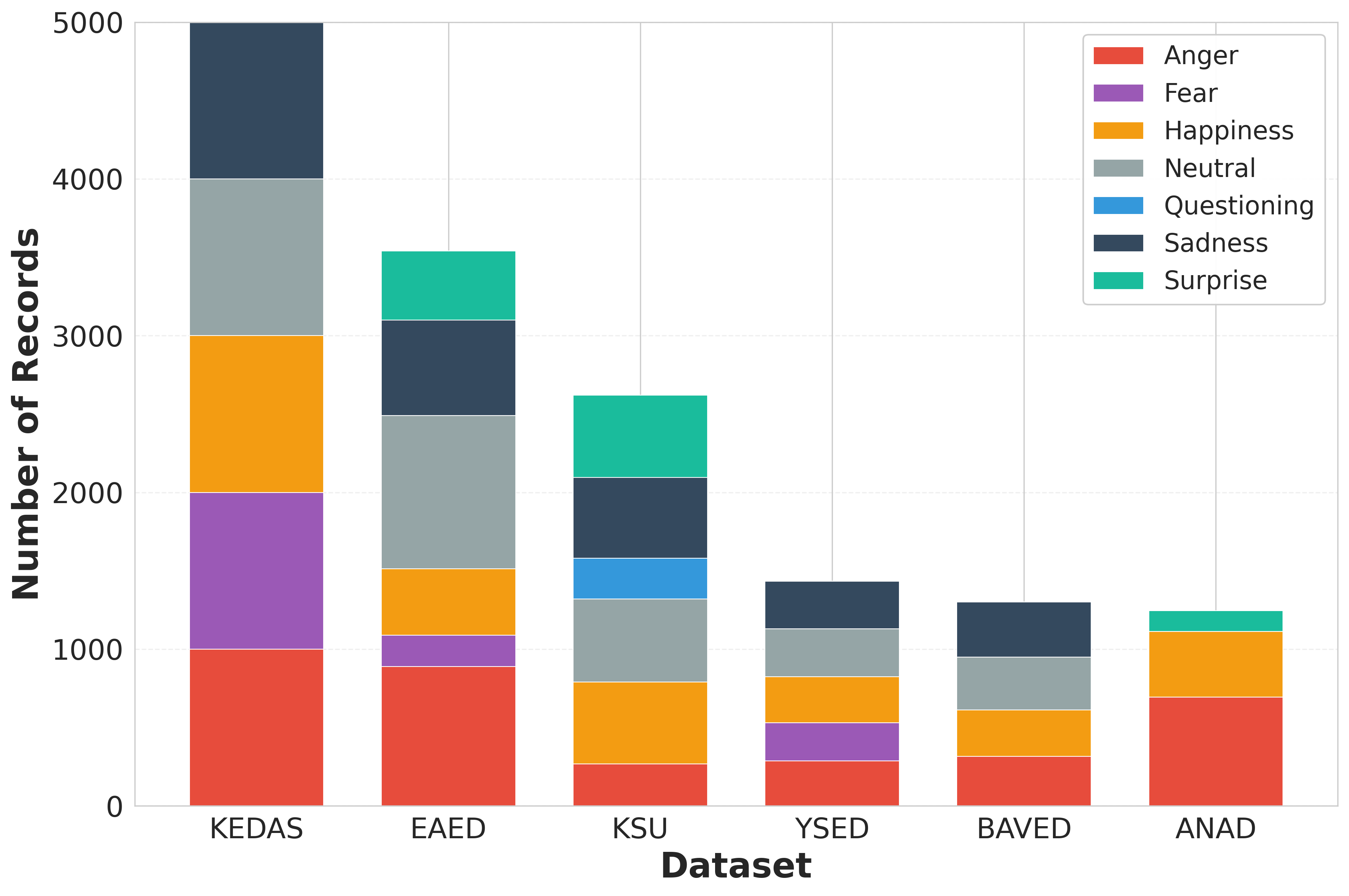}
    \vspace{-0.2cm}
    \caption{Emotion label distribution across training datasets.}
    \label{fig:emotion_train_distribution}
    \vspace{0.2cm}
\end{figure}

\begin{figure}[h]
    \centering
    \includegraphics[width=0.95\linewidth]{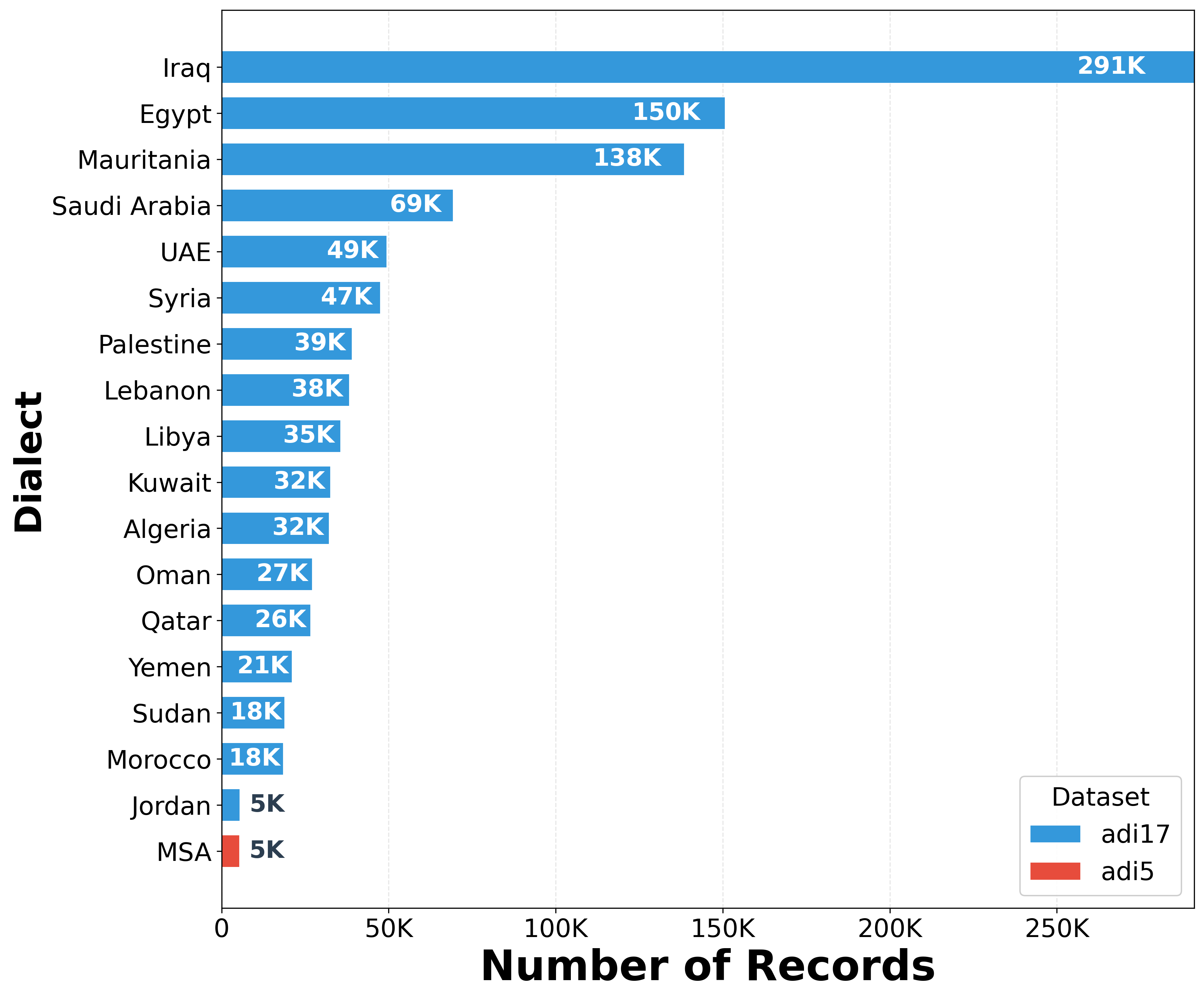}
    \vspace{0.3cm}
    \caption{Dialect distribution in the training set.}
    \label{fig:did_disi}
    \vspace{0.3cm}
\end{figure}

\begin{figure}[!t]
    \centering
    \includegraphics[width=0.8\linewidth]{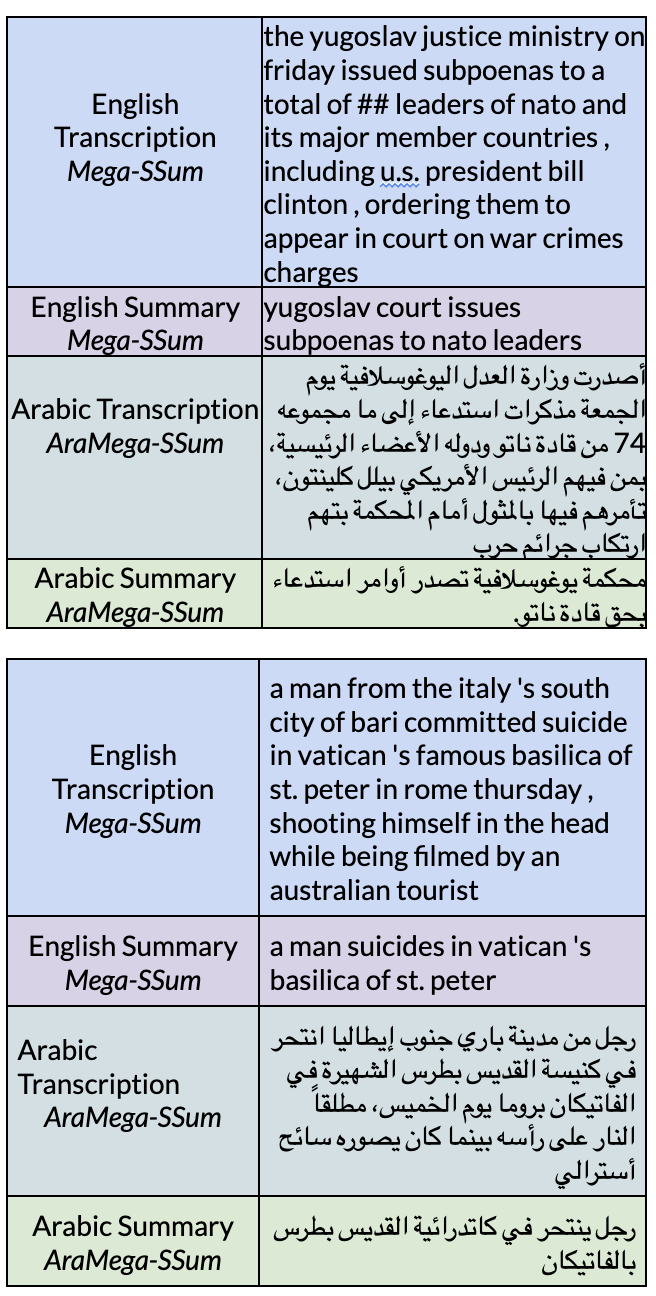}
    \vspace{-0.2cm}
    \caption{Sample speech summarization instances from AraMega-SSum.}
    \label{fig:ssum_eg}
    \vspace{-0.3cm}
\end{figure}

Figure~\ref{fig:emotion_train_distribution} presents the distribution of emotion labels across multiple Arabic emotion datasets used for training, including KEDAS, EAED, KSU, YSED, BAVED, and ANAD. Each dataset exhibits a distinct class composition, with emotions such as anger, happiness, neutrality, and sadness consistently represented, while fear, questioning, and surprise appear less frequently and vary across datasets. KEDAS and EAED contribute the largest number of samples with a broad coverage of emotion categories.

Figure~\ref{fig:did_disi} shows the training data covers dialects across 17 countries using the \textbf{ADI-17} dataset, along with MSA from \textbf{ADI-5}. The distribution is dominated by Egyptian, Iraqi, and Mauritanian dialects, while the remaining dialects are represented with varying but comparatively smaller sample sizes.

Figure~\ref{fig:ssum_eg} presents example samples from the MegaSSUM dataset, showing English transcriptions and summaries, alongside corresponding Arabic transcriptions and summaries from the proposed AraMega-SSum dataset. Each instance is paired with both English and Arabic speech through text-to-speech generation, enabling cross-lingual and cross-modal speech summarization training and evaluation.

\section{AraMega-SSum Validation}
\label{sec_app_aramega_ssum_validation}

AraMega-SSum is the only newly developed semi-synthetic dataset introduced in this work. The other datasets used for ASR, emotion recognition, and dialect identification are curated from publicly available real-speech benchmarks. We therefore provide additional validation for AraMega-SSum to clarify its construction process, intended use, and quality-control procedures.

AraMega-SSum is designed to support Arabic speech summarization, a setting in which publicly available datasets with aligned speech, transcripts, and abstractive summaries remain scarce. The dataset serves as a quality-controlled semi-synthetic resource that enables controlled experimentation in this under-resourced setting.

It is derived from MegaSSum. We translate the source transcripts and abstractive summaries into Arabic using Gemini-2.5-Flash, followed by manual checks to improve semantic consistency, information preservation, contextual accuracy, completeness, and coherence. For speech generation, we use neural voice cloning conditioned on real speaker references. The generated speech is then evaluated through both human assessment and automatic quality checks.

\subsection{Translation Quality}

In Table~\ref{tab:translation_quality}, we present English-to-Arabic summary translations achieve near-perfect evaluation scores from a GPT-4.1 judge, consistently exceeding 9.95 out of 10 across all measured quality metrics, including semantic equivalence and coherence. On the human evaluation we observe similar scores in all rubrics as shown in Table~\ref{tab:translation_quality_hh}.

\begin{table}[!tbh]
\centering
\scalebox{0.9}{
\small
\setlength{\tabcolsep}{6pt}
\renewcommand{\arraystretch}{1.05}
\begin{tabular}{l c}
\toprule
\textbf{Metric (EN$\rightarrow$AR transfer)} & \textbf{Score (/10)} \\
\midrule
Semantic Equivalence      & 9.96 \\
Information Preservation  & 9.96 \\
Contextual Accuracy       & 9.96 \\
Completeness              & 9.96 \\
Coherence                 & 9.98 \\
\bottomrule
\end{tabular}}
\vspace{-0.3cm}
\caption{LLM-as-a-judge (GPT-4.1) scores for English$\rightarrow$Arabic summary translation by Gemini. Scores are out of 10.}
\label{tab:translation_quality}
\vspace{-0.3cm}
\end{table}

\begin{table}[!tbh]
\centering
\scalebox{0.9}{
\small
\setlength{\tabcolsep}{6pt}
\renewcommand{\arraystretch}{1.05}
\begin{tabular}{l c}
\toprule
\textbf{Metric (EN$\rightarrow$AR transfer)} & \textbf{Score (/10)} \\
\midrule
Semantic Equivalence      & 9.93 \\
Information Preservation  & 9.95 \\
Contextual Accuracy       & 9.99 \\
Completeness              & 9.95 \\
Coherence                 & 10.0 \\
\bottomrule
\end{tabular}}
\vspace{-0.3cm}
\caption{Human scores for English$\rightarrow$Arabic translation quality. Scores are out of 10.}
\label{tab:translation_quality_hh}
\vspace{-0.3cm}
\end{table}

\subsection{Audio Quality}

For speech generation, we use neural voice cloning conditioned on real speaker references. To validate the generated Arabic speech, we conduct both human and automatic quality checks. For the human evaluation, three paid annotators evaluate 100 randomly sampled audio clips from the test set. The annotators rate each clip on a 5-point Likert scale, where 5 means excellent, 4 good, 3 fair, 2 poor, and 1 bad. We evaluate three standard speech-quality dimensions: naturalness, audio quality, and intelligibility. As shown in Table~\ref{tab:human-speech-quality}, the generated speech receives good average scores across the three dimensions. We also report Gwet's AC1~\cite{gwet2008computing} to measure annotator agreement, which shows substantial agreement for naturalness and intelligibility and very high agreement for audio quality.

\begin{table}[!tbh]
\centering
\small
\setlength{\tabcolsep}{6pt}
\begin{tabular}{lcc}
\toprule
\textbf{Metric} & \textbf{Avg.} & \textbf{Gwet's AC1} \\
\midrule
Naturalness    & 3.975 & 78.9\% \\
Audio Quality  & 4.373 & 95.0\% \\
Intelligibility & 3.692 & 78.9\% \\
\bottomrule
\end{tabular}
\vspace{-0.3cm}
\caption{Human speech-quality evaluation of 100 randomly sampled AraMega-SSum test-set audio clips.}
\label{tab:human-speech-quality}
\vspace{-0.3cm}
\end{table}

In addition to the human evaluation, we report automatic speech-quality and speaker-consistency indicators. Specifically, we compute ASR word error rate using Fanar Aura ASR, overall speech quality using NISQA, and speaker cosine similarity between the generated audio and the corresponding speaker reference. The generated speech obtains a Fanar Aura ASR WER of $15.5$, a NISQA overall score of $3.95$ out of 5, and a speaker cosine similarity of $0.664$. These values fall within acceptable ranges and are consistent with prior studies~\cite{alam2025everydaymmqa}.

Overall, these results show that AraMega-SSum is a quality-controlled semi-synthetic resource for Arabic speech summarization. The dataset fills an important resource gap while keeping its semi-synthetic nature explicit. We therefore use it as a controlled benchmark for Arabic speech summarization experiments.

\section{Prompts}
\label{sec_app_prompts}

In Figures~\ref{fig:llm_judge_prompt} and~\ref{fig:llm_translation_judge_prompt} and Listing~\ref{lst:arabic-audio-prompts}, we present the prompts used for LLM-as-judge summarization evaluation, LLM-as-judge translation evaluation, and multitask training.

\lstdefinelanguage{json}{
  basicstyle=\ttfamily\footnotesize,
  numbers=none,
  numberstyle=\tiny,
  stepnumber=1,
  numbersep=6pt,
  showstringspaces=false,
  breaklines=true,
  frame=single,
  backgroundcolor=\color{gray!5},
  stringstyle=\color{orange!80!black},
  identifierstyle=\color{blue!70!black},
  keywordstyle=\color{blue!70!black},
  morestring=[b]",
  literate=
   *{0}{{{\color{violet!70!black}0}}}{1}
    {1}{{{\color{violet!70!black}1}}}{1}
    {2}{{{\color{violet!70!black}2}}}{1}
    {3}{{{\color{violet!70!black}3}}}{1}
    {4}{{{\color{violet!70!black}4}}}{1}
    {5}{{{\color{violet!70!black}5}}}{1}
    {6}{{{\color{violet!70!black}6}}}{1}
    {7}{{{\color{violet!70!black}7}}}{1}
    {8}{{{\color{violet!70!black}8}}}{1}
    {9}{{{\color{violet!70!black}9}}}{1}
    {:}{{{\color{black}:}}}{1}
    {,}{{{\color{black},}}}{1}
    {\{}{{{\color{black}\{}}}{1}
    {\}}{{{\color{black}\}}}}{1}
    {[}{{{\color{black}[}}}{1}
    {]}{{{\color{black}]}}}{1}
}

\begin{figure}[t]
\centering
\lstset{
    basicstyle=\ttfamily\footnotesize,
    breaklines=true,
    breakatwhitespace=true,
    columns=fullflexible,
    frame=single,
    keepspaces=true,
    showstringspaces=false
}
\begin{lstlisting}[language=Python]
SYSTEM_PROMPT = """You are a reference-grounded summarization quality evaluator.
Grade a predicted summary by comparing it with the reference summary.
Do NOT use outside knowledge. Judge in the same language as the summaries
(Arabic or English). Return ONLY a valid JSON object that matches the
schema exactly—no extra text.

Score each criterion as an INTEGER from 1 to 10 (1=poor, 10=excellent):

- Clarity
- Conciseness
- Coherence
- Completeness
- Semantic_Alignment
- Accuracy
- Relevance
- Information_Density

"""

USER_PROMPT_TEMPLATE = """Language: {language}
Reference summary: {reference_summary}
Predicted summary: {predicted_summary}

Evaluate the predicted summary by comparing it with the reference summary.
Output only the JSON described in the system prompt."""
\end{lstlisting}
\caption{\textbf{Prompt used for LLM-as-a-judge summarization evaluation.}}
\label{fig:llm_judge_prompt}
\end{figure}

\begin{figure*}[!tbh]
\centering
\lstset{
    basicstyle=\ttfamily\footnotesize,
    breaklines=true,
    breakatwhitespace=true,
    columns=fullflexible,
    frame=single,
    keepspaces=true,
    showstringspaces=false
}
\begin{minipage}{\textwidth}
\begin{lstlisting}[language=Python]
SYSTEM_PROMPT = """You are an expert translation evaluator.

You will be given:
- An English transcription (may contain anonymized tokens such as ###, ####,
  and other placeholders)
- An Arabic transcription translated from the same English content

Your task is to evaluate how semantically equivalent the two transcriptions are.

Evaluation Rules (STRICT):
1. Ignore all anonymization tokens in English.
2. Ignore number differences caused by anonymization.
3. Ignore name differences and transliteration variations.
4. Be lenient with phonetic spellings in Arabic.
5. Focus only on semantic meaning and core information.
6. Judge based on events, facts, actions, relationships, and intent.

Score each criterion as an INTEGER from 1 to 10:

- semantic_equivalence
- information_preservation
- contextual_accuracy
- completeness
- coherence

Return ONLY a valid JSON object matching the schema exactly.
"""

USER_PROMPT_TEMPLATE = """Arabic translation:
{arabic_transcription}

Original English source:
{english_transcription}

Evaluate the quality of the Arabic translation relative to the English source.
Follow the evaluation rules defined in the system prompt.
Return only the required JSON output."""
\end{lstlisting}
\end{minipage}
\caption{\textbf{Prompt used for LLM-as-a-judge translation evaluation.}}
\label{fig:llm_translation_judge_prompt}
\end{figure*}

\begin{figure*}[!tbh]
\centering
\begin{minipage}{0.98\textwidth}
\begin{lstlisting}[
  language=prompt,
  basicstyle=\ttfamily\footnotesize,
  breaklines=true,
  columns=fullflexible,
  frame=single,
]
System
You are an Arabic audio-language assistant. You will receive audio and a user instruction describing the task. Your capabilities include: (1) Accurate speech transcription with proper formatting, (2) Dialect and emotion identification from speech, (3) Understanding conversational context and dialogue acts, (4) Generating concise, coherent summaries from both speech and text. Always follow instructions precisely, maintain linguistic accuracy, and format outputs exactly as requested. For ASR/transcription tasks, output the spoken words verbatim in Arabic as written, preserving numbers, names, and code-switching without paraphrasing or summarizing.

ASR
<audio>
Task: Transcription.
Transcribe the audio accurately in its original language.
Respond with a single-line JSON object only:
{"transcription":"<text>"}

Dialect
<audio>
Task: Dialect Identification.
Identify the dialect spoken in the audio.
Choose exactly one label from the following list:
Algeria, Egypt, Iraq, Jordan, Kuwait, Lebanon, Libya, Mauritania, Modern Standard Arabic, Morocco, Oman, Palestine, Qatar, Saudi Arabia, Sudan, Syria, United Arab Emirates, Yemen

Respond with a single-line JSON object only:
{"dialect":"<dialect>"}

Emotion
<audio>
Task: Emotion Identification.
Identify the primary emotion expressed in the audio.
Choose exactly one emotion from the following list:
Anger, Fear, Happiness, Neutral, Questioning, Sadness, Surprise

Respond with a single-line JSON object only:
{"emotion":"<emotion>"}

SSUM
<audio>
Task: Speech Summarization.
Summarize the main content of the audio concisely.
Preserve the original language of the speech.

Respond with a single-line JSON object only:
{"summary":"<text>"}

TSUM
Task: Text Summarization.
Read the following text: "{text}"

Summarize it concisely in the same language.
Respond with a single-line JSON object only:
{"summary":"<text>"}
\end{lstlisting}
\end{minipage}
  \caption{\textbf{System and user prompts used for multi-task training.}}
  \label{lst:arabic-audio-prompts}
\end{figure*}

\section{Training Loss}
\label{sec_app_training_loss}

\begin{figure}[t]
  \centering
  \includegraphics[width=\linewidth]{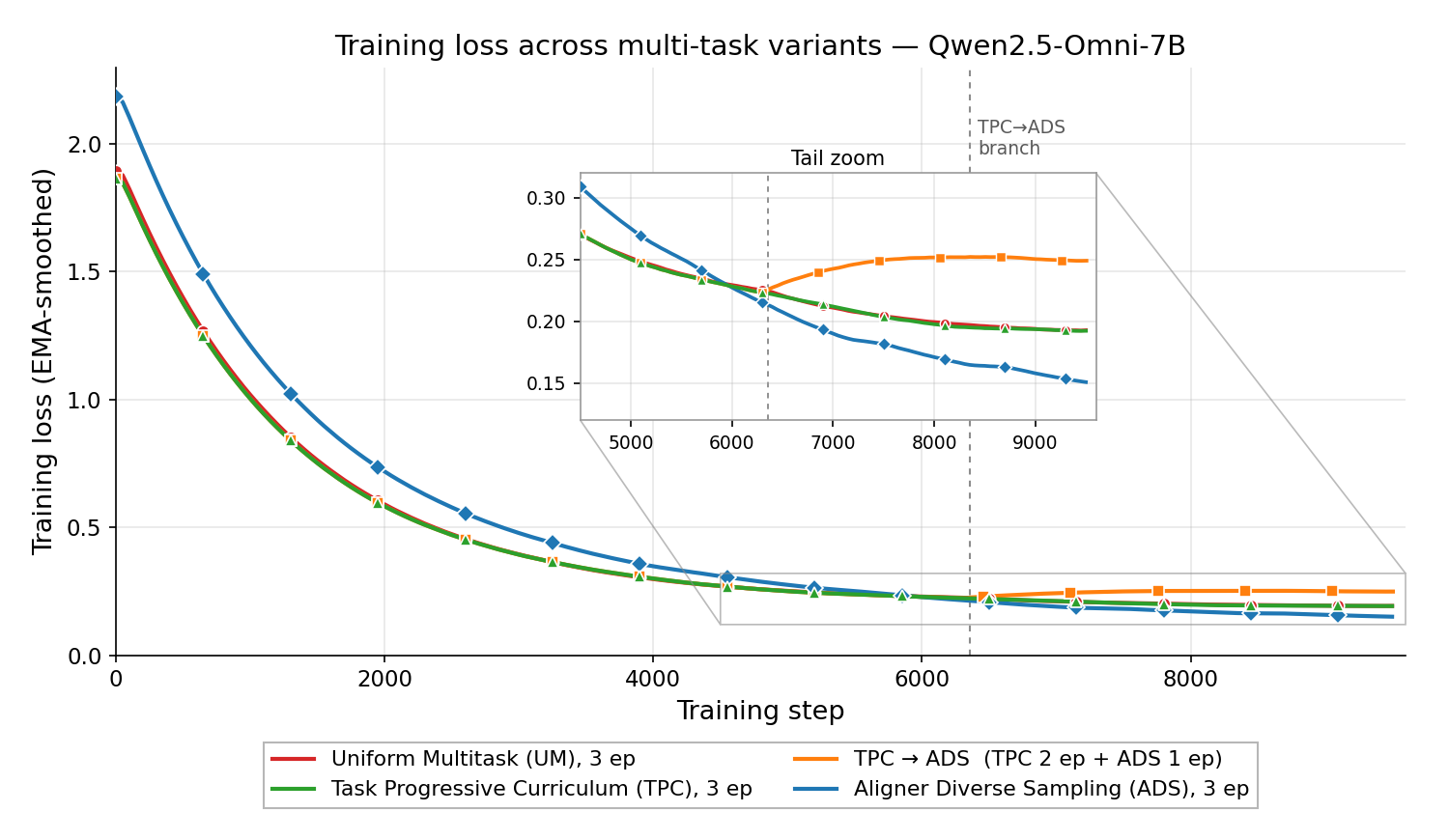}
  \caption{Training loss across multi-task variants for Qwen2.5-Omni-7B.}
  \label{fig:loss-qwen}
\end{figure}

\begin{figure}[t]
  \centering
  \includegraphics[width=\linewidth]{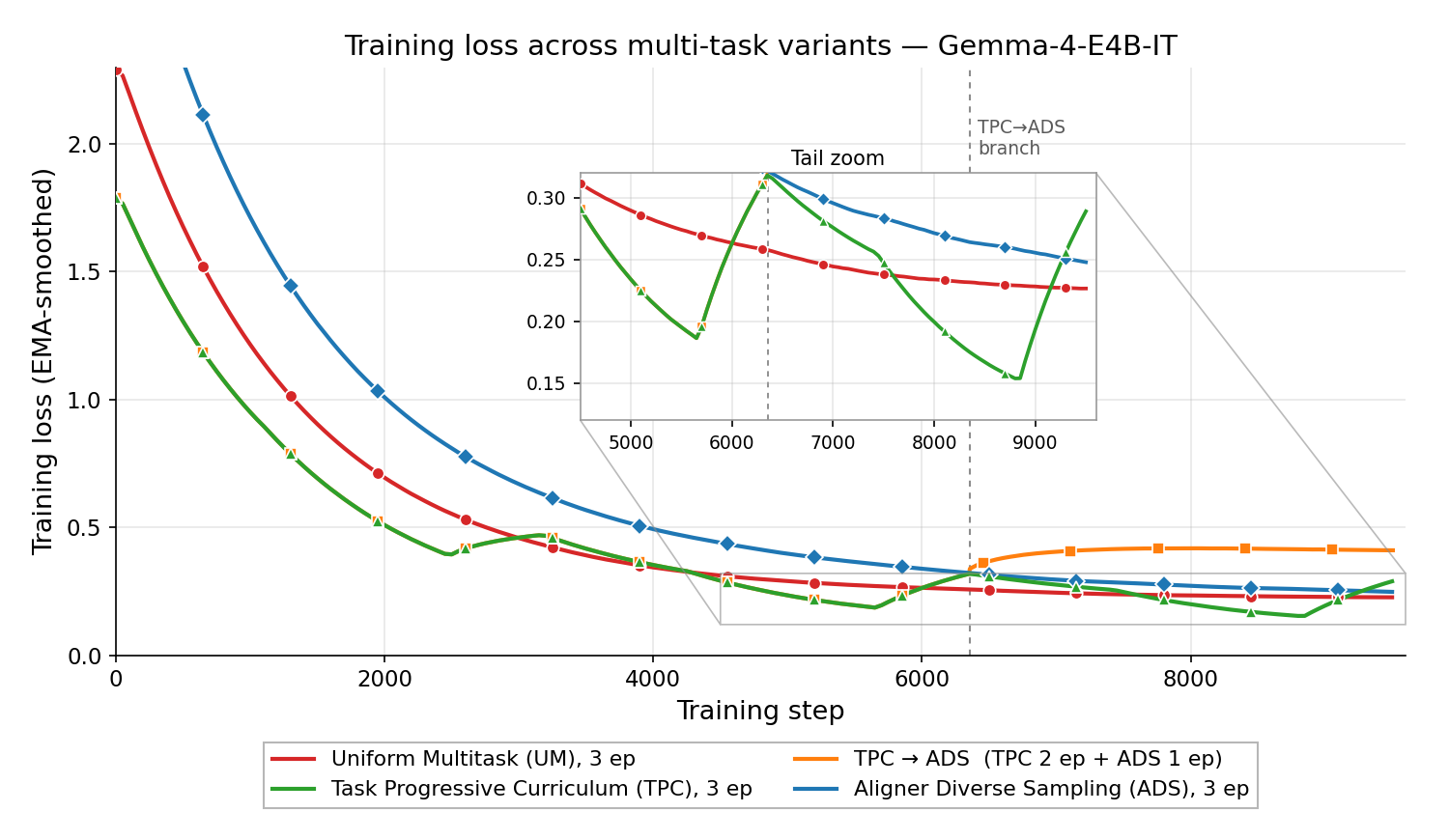}
  \caption{Training loss across multi-task variants for Gemma-4-E4B-IT.}
  \label{fig:loss-gemma}
\end{figure}

Figures~\ref{fig:loss-qwen} and~\ref{fig:loss-gemma} show the smoothed training-loss curves for the four multi-task fine-tunes, which all start from the same ASR-aligned LoRA checkpoint and differ only in how the multi-task data is presented. Uniform Multitask (UM) randomly shuffles all tasks together. Task Progressive Curriculum (TPC) reorders the same data so that each epoch progresses from easier to harder tasks, restarting the curriculum every epoch---visible as the intra-epoch U-shapes on Gemma TPC. Aligner Diverse Sampling (ADS), shown in blue, weights samples by aligner diversity, producing a higher early loss but the lowest final training loss on both backbones. TPC $\rightarrow$ ADS is our proposed hybrid: two epochs of curriculum to build a structured representation, followed by a third epoch of diversity-weighted sampling to broaden coverage. The step-6356 transition appears as a visible uptick where the distribution shifts, after which the loss re-stabilizes. The tail-zoom insets highlight that UM and TPC converge to comparable final loss (approximately 0.19 on Qwen and 0.23 on Gemma), confirming that the curriculum remains competitive with random shuffling at matched compute.

\section{Additional Results}
\label{sec_app_additional_results}

In Table~\ref{tab:asr-detailed-wer}, we provide a detailed ASR breakdown across language varieties and speech conditions. The results show that adaptation consistently improves over the base Qwen2.5-Omni-7B model across MSA, dialectal Arabic, code-switched speech, native English, and L2 English. LA gives large gains over the base model, especially for English ASR, but the multi-task phases further improve Arabic and code-switched speech. For Qwen2.5-Omni-7B, TPC and TPC$\rightarrow$ADS achieve the strongest Arabic and code-switching results, while UM remains highly competitive. For Gemma-4-E4B-it, UM and TPC$\rightarrow$ADS perform best across most Arabic and code-switched settings, with TPC$\rightarrow$ADS giving the lowest WER on several dialectal and code-switched benchmarks. The closed models perform strongly on some subsets, but the adapted open models remain competitive, especially on MSA, English, and several dialectal or code-switched conditions. Overall, the detailed breakdown supports the aggregate ASR trends in the main results and shows that adaptation improves robustness across diverse speech conditions.

\begin{table*}[t]
\centering
\setlength{\tabcolsep}{3.2pt}
\scalebox{0.7}{
\begin{tabular}{llrrrrrrrr}
\toprule
\textbf{Model} & \textbf{Phase}
& \multicolumn{1}{c}{\textbf{MSA}}
& \multicolumn{2}{c}{\textbf{DA}}
& \multicolumn{2}{c}{\textbf{CS}}
& \multicolumn{2}{c}{\textbf{En L1}}
& \multicolumn{1}{c}{\textbf{En L2}} \\
\cmidrule(lr){3-3}
\cmidrule(lr){4-5}
\cmidrule(lr){6-7}
\cmidrule(lr){8-9}
\cmidrule(lr){10-10}
& & \textbf{MGB2} & \textbf{MGB3} & \textbf{SADA} & \textbf{ESCWA} & \textbf{DACS} & \textbf{Libri-clean} & \textbf{Libri-other} & \textbf{L2-Arctic} \\
\midrule
Qwen2.5-Omni-7B & Base & 33.48 & 54.59 & 76.62 & 61.91 & 45.53 & 41.51 & 39.20 & 37.47 \\
Qwen2.5-Omni-7B & LA & 13.51 & 31.53 & 54.87 & 42.53 & 22.72 & 1.46 & 2.96 & 4.02 \\
Qwen2.5-Omni-7B & UM & 12.98 & 27.49 & 43.61 & 40.70 & 20.09 & 1.48 & 2.82 & 2.90 \\
Qwen2.5-Omni-7B & TPC & 12.98 & 27.46 & 43.66 & 40.31 & 19.94 & 1.45 & 2.86 & 3.10 \\
Qwen2.5-Omni-7B & ADS & 13.50 & 27.98 & 44.02 & 43.32 & 20.18 & 1.47 & 2.93 & 3.39 \\
Qwen2.5-Omni-7B & TPC$\rightarrow$ADS & 13.21 & 27.28 & 43.79 & 40.64 & 20.06 & 1.48 & 2.90 & 2.95 \\
\midrule
Gemma-4-E4B-it & Base & 16.70 & 30.97 & 42.60 & 43.81 & 22.70 & 3.86 & 9.34 & 13.30 \\
Gemma-4-E4B-it & LA & 13.77 & 26.87 & 44.32 & 34.37 & 19.79 & 2.57 & 6.59 & 12.64 \\
Gemma-4-E4B-it & UM & 12.42 & 24.05 & 40.06 & 41.95 & 17.57 & 2.27 & 5.85 & 7.31 \\
Gemma-4-E4B-it & TPC & 12.88 & 23.95 & 40.27 & 43.15 & 17.98 & 2.27 & 5.90 & 7.74 \\
Gemma-4-E4B-it & ADS & 13.36 & 24.48 & 43.39 & 40.90 & 18.52 & 2.58 & 6.22 & 8.53 \\
Gemma-4-E4B-it & TPC$\rightarrow$ADS & 13.00 & 23.81 & 39.65 & 39.99 & 17.97 & 2.28 & 5.99 & 10.45 \\
\midrule
Qwen2.5-Omni-3B & Base & 45.79 & 62.43 & 85.41 & 71.50 & 57.16 & 33.74 & 35.45 & 48.13 \\
GPT-audio & Base & 14.44 & 30.38 & 46.88 & 33.52 & 22.97 & 7.16 & 10.70 & 15.25 \\
Gemini-2.5-Pro & Base & 18.98 & 26.55 & 34.91 & 38.92 & 20.18 & 3.70 & 6.84 & 8.70 \\
\bottomrule
\end{tabular}
}
\caption{
Detailed ASR WER results across language varieties and speech conditions.
We report WER for MSA, dialectal Arabic, Arabic--English code-switching, dialect--MSA code-switching, native English, and L2 English.
Lower scores indicate better performance.
}
\label{tab:asr-detailed-wer}
\end{table*}

Table~\ref{tab:ssum-tsum-detailed} reports detailed SSUM and TSUM results across English and Arabic using ROUGE-1, ROUGE-L, and BERTScore. Multi-task adaptation improves both Qwen2.5-Omni-7B and Gemma-4-E4B-it over their base and LA settings across most metrics. For Qwen2.5-Omni-7B, UM and TPC achieve the strongest SSUM results, while TPC and TPC$\rightarrow$ADS perform best on TSUM. For Gemma-4-E4B-it, UM gives the best SSUM scores, whereas TPC and TPC$\rightarrow$ADS lead on TSUM. These results support the main finding that UM and TPC provide strong summarization performance, while TPC$\rightarrow$ADS gives targeted gains in selected TSUM settings.

\begin{table*}[t]
\centering
\small
\setlength{\tabcolsep}{3.2pt}
\scalebox{0.75}{
\begin{tabular}{llrrrrrrrrrrrr}
\toprule
\textbf{Model} & \textbf{Phase}
& \multicolumn{6}{c}{\textbf{SSUM}}
& \multicolumn{6}{c}{\textbf{TSUM}} \\
\cmidrule(lr){3-8}
\cmidrule(lr){9-14}
& &
\multicolumn{3}{c}{\textbf{English}} &
\multicolumn{3}{c}{\textbf{Arabic}} &
\multicolumn{3}{c}{\textbf{English}} &
\multicolumn{3}{c}{\textbf{Arabic}} \\
\cmidrule(lr){3-5}
\cmidrule(lr){6-8}
\cmidrule(lr){9-11}
\cmidrule(lr){12-14}
& &
\textbf{R-1}$\uparrow$ & \textbf{R-L}$\uparrow$ & \textbf{B-F1}$\uparrow$ &
\textbf{R-1}$\uparrow$ & \textbf{R-L}$\uparrow$ & \textbf{B-F1}$\uparrow$ &
\textbf{R-1}$\uparrow$ & \textbf{R-L}$\uparrow$ & \textbf{B-F1}$\uparrow$ &
\textbf{R-1}$\uparrow$ & \textbf{R-L}$\uparrow$ & \textbf{B-F1}$\uparrow$ \\
\midrule
Qwen2.5-Omni-7B & Base             & 29.37 & 26.48 & 57.37 & 23.70 & 21.53 & 53.75 & 25.67 & 23.30 & 51.13 & 29.84 & 27.56 & 57.86 \\
Qwen2.5-Omni-7B & LA               & 25.75 & 23.05 & 56.61 & 23.84 & 21.69 & 53.59 & 24.35 & 21.87 & 52.87 & 25.34 & 23.11 & 54.22 \\
Qwen2.5-Omni-7B & UM               & 46.23 & \textbf{43.62} & \textbf{69.79} & 36.74 & 34.71 & 63.14 & 48.14 & 45.41 & 70.78 & 38.20 & 36.06 & \textbf{63.97} \\
Qwen2.5-Omni-7B & TPC              & \textbf{46.36} & 43.61 & 69.75 & \textbf{36.75} & \textbf{34.74} & \textbf{63.21} & \textbf{48.26} & \textbf{45.55} & \textbf{70.85} & 38.16 & 36.09 & 63.89 \\
Qwen2.5-Omni-7B & ADS              & 41.43 & 38.79 & 67.32 & 33.66 & 31.65 & 61.23 & 43.56 & 40.86 & 68.41 & 35.77 & 33.66 & 62.52 \\
Qwen2.5-Omni-7B & TPC$\rightarrow$ADS & 45.16 & 42.40 & 69.20 & 36.14 & 34.14 & 62.92 & 47.78 & 45.13 & 70.60 & \textbf{38.28} & \textbf{36.16} & 63.85 \\
\midrule
Gemma-4-E4B-it  & Base             & 25.98 & 23.20 & 56.02 & 24.39 & 22.21 & 54.04 & 23.19 & 20.78 & 49.47 & 29.75 & 27.35 & 57.51 \\
Gemma-4-E4B-it  & LA               & 20.53 & 18.32 & 50.31 & 24.00 & 21.85 & 53.70 &  1.86 &  1.71 & 25.60 & 26.24 & 23.96 & 55.02 \\
Gemma-4-E4B-it  & UM               & \textbf{45.32} & \textbf{42.57} & \textbf{69.05} & \textbf{37.49} & \textbf{35.46} & \textbf{63.53} & 49.46 & 46.71 & 71.19 & 39.89 & 37.76 & 64.80 \\
Gemma-4-E4B-it  & TPC              & 45.03 & 42.31 & 68.87 & 36.76 & 34.81 & 63.37 & \textbf{50.09} & \textbf{47.20} & \textbf{71.46} & 39.76 & 37.69 & 64.90 \\
Gemma-4-E4B-it  & ADS              & 42.48 & 40.01 & 67.71 & 34.46 & 32.55 & 61.97 & 48.22 & 45.30 & 70.52 & 38.01 & 35.78 & 63.70 \\
Gemma-4-E4B-it  & TPC$\rightarrow$ADS & 43.98 & 41.24 & 68.34 & 36.34 & 34.27 & 63.07 & 49.78 & 46.81 & 71.30 & \textbf{40.04} & \textbf{37.99} & \textbf{64.99} \\
\midrule
Qwen2.5-Omni-3B & Base             & 13.13 & 11.79 & 38.90 & 22.05 & 19.95 & 52.49 & 14.36 & 12.95 & 38.33 & 27.68 & 25.31 & 56.34 \\
GPT-audio       & Base             &  5.48 &  5.21 & 29.62 & 28.63 & 26.24 & 57.42 & --    & --    & --    & --    & --    & --    \\
GPT-5-chat      & Base             & --    & --    & --    & --    & --    & --    & 34.08 & 30.87 & 60.83 & 31.07 & 28.56 & 59.41 \\
Gemini-2.5-Pro  & Base             & 28.48 & 25.75 & 57.01 & 27.70 & 25.40 & 56.15 & 33.20 & 30.35 & 59.74 & 34.13 & 31.59 & 60.55 \\
\bottomrule
\end{tabular}
}
\caption{
Detailed summarization results for SSUM and TSUM across English and Arabic. We report ROUGE-1, ROUGE-L, and BERTScore F1 (B-F1). Higher scores indicate better performance. Bold marks the best score within each adapted open-model block.
}
\label{tab:ssum-tsum-detailed}
\end{table*}

Table~\ref{tab:ssum-llm-judge} reports LLM-as-a-judge scores for SSUM across English and Arabic. The judge scores are broadly consistent with the automatic metrics: UM and TPC remain competitive across both models, ADS scores lowest among adapted phases, and Gemini-2.5-Pro leads overall.

\begin{table}[t]
\centering
\small
\setlength{\tabcolsep}{5pt}
\begin{tabular}{llrr}
\toprule
\textbf{Model} & \textbf{Phase} & \textbf{EN}$\uparrow$ & \textbf{AR}$\uparrow$ \\
\midrule
Qwen2.5-Omni-7B & Base             & 8.82 & 7.26 \\
Qwen2.5-Omni-7B & LA               & 6.87 & 7.55 \\
Qwen2.5-Omni-7B & UM               & 8.66 & 7.66 \\
Qwen2.5-Omni-7B & TPC              & 8.63 & 7.65 \\
Qwen2.5-Omni-7B & ADS              & 8.13 & 7.23 \\
Qwen2.5-Omni-7B & TPC$\rightarrow$ADS & 8.58 & 7.61 \\
\midrule
Gemma-4-E4B-it  & Base             & 8.13 & 7.95 \\
Gemma-4-E4B-it  & LA               & 6.45 & 7.79 \\
Gemma-4-E4B-it  & UM               & \textbf{8.37} & 7.71 \\
Gemma-4-E4B-it  & TPC              & 8.31 & \textbf{7.74} \\
Gemma-4-E4B-it  & ADS              & 8.11 & 7.34 \\
Gemma-4-E4B-it  & TPC$\rightarrow$ADS & 8.29 & 7.65 \\
\midrule
Qwen2.5-Omni-3B & Base             & 6.32 & 6.59 \\
GPT-audio       & Base             & 6.60 & 8.97 \\
Gemini-2.5-Pro  & Base             & \textbf{9.03} & \textbf{9.01} \\
\bottomrule
\end{tabular}
\caption{LLM-as-a-judge scores for SSUM across English and Arabic (out of 10). Higher is better. Bold marks the best score within each adapted open-model block and among baselines respectively.}
\label{tab:ssum-llm-judge}
\end{table}

Table~\ref{tab:did-detailed} reports detailed dialect identification results on ADI17. Multi-task adaptation strongly improves both open models over their base and LA settings. For Qwen2.5-Omni-7B, TPC$\rightarrow$ADS achieves the best accuracy, macro-F1, and weighted-F1, while TPC gives the second-best macro-F1. For Gemma-4-E4B-it, TPC$\rightarrow$ADS achieves the best accuracy and weighted-F1, while UM gives the best macro-F1. These results show that adaptation is especially important for dialect identification. They also support the main finding that TPC$\rightarrow$ADS is highly effective for discriminative speech understanding, although the strongest phase can still vary by model and metric.

\begin{table}[t]
\centering
\small
\setlength{\tabcolsep}{4.5pt}
\begin{tabular}{llrrr}
\toprule
\textbf{Model} & \textbf{Phase}
& \textbf{Acc.} $\uparrow$
& \textbf{M-F1} $\uparrow$
& \textbf{W-F1} $\uparrow$ \\
\midrule
Qwen2.5-Omni-7B & Base & 13.77 & 7.56 & 10.31 \\
Qwen2.5-Omni-7B & LA & 5.08 & 0.01 & 8.58 \\
Qwen2.5-Omni-7B & UM & 87.91 & 82.80 & 87.76 \\
Qwen2.5-Omni-7B & TPC & \underline{87.93} & \underline{87.71} & 87.78 \\
Qwen2.5-Omni-7B & ADS & 86.77 & 77.49 & 86.72 \\
Qwen2.5-Omni-7B & TPC$\rightarrow$ADS & \textbf{88.85} & \textbf{88.80} & \textbf{88.80} \\
\midrule
Gemma-4-E4B-it & Base & 26.62 & 17.76 & 21.17 \\
Gemma-4-E4B-it & LA & 3.31 & 0.01 & 6.17 \\
Gemma-4-E4B-it & UM & \underline{79.28} & \textbf{78.89} & 78.83 \\
Gemma-4-E4B-it & TPC & 79.03 & \underline{78.57} & 78.55 \\
Gemma-4-E4B-it & ADS & 76.03 & 71.79 & 75.93 \\
Gemma-4-E4B-it & TPC$\rightarrow$ADS & \textbf{80.12} & 75.68 & \textbf{79.99} \\
\midrule
Qwen2.5-Omni-3B & Base & 11.14 & 7.14 & 7.66 \\
GPT-audio & Base & 40.96 & 25.04 & 38.44 \\
Gemini-2.5-Pro & Base & 65.09 & 59.85 & 63.37 \\
\bottomrule
\end{tabular}
\caption{
Detailed dialect identification results on ADI17. We report accuracy, macro-F1, and weighted-F1. Higher scores indicate better performance. Bold marks the best score within each adapted open-model block, and underlining marks the second best.
}
\label{tab:did-detailed}
\end{table}

\end{document}